\documentclass[twocolumn]{aastex631}
\usepackage{graphicx}   
\usepackage{lipsum} 
\usepackage{amsmath}
\usepackage{appendix}

\usepackage{color}

\def\hmsun{{h^{-1} M_{\odot}}}

\def\astrid{\texttt{ASTRID} }

\defcitealias{NG15-binary}{NG15-binary}
\defcitealias{NG15-evidence}{NG15-gwb}

\shorttitle{ASTRID GWB}
\shortauthors{N.Chen et al.}
\graphicspath{{./}{figures/}}

\begin{document}


\title{The Gravitational Wave Background from Massive Black Holes in the ASTRID Simulation}

\author{Nianyi Chen}
\affiliation{School of Natural Sciences, Institute for Advanced Study, Princeton, NJ 08540, USA}
\affiliation{McWilliams Center for Cosmology, Department of Physics, Carnegie Mellon University, Pittsburgh, PA 15213, USA}

\author{Tiziana Di Matteo}
\affiliation{McWilliams Center for Cosmology, Department of Physics, Carnegie Mellon University, Pittsburgh, PA 15213, USA}

\author{Yihao Zhou}
\affiliation{McWilliams Center for Cosmology, Department of Physics, Carnegie Mellon University, Pittsburgh, PA 15213, USA}

\author {Luke Zoltan Kelley}
\affiliation{Department of Astronomy, University of California, Berkeley, Berkeley, CA 94720, USA}

\author {Laura Blecha}
\affiliation{Physics Department, University of Florida, Gainesville, FL 32611, USA}

\author{Yueying Ni}
\affiliation{Center for Astrophysics $\vert$ Harvard \& Smithsonian, Cambridge, MA 02138, USA}

\author{Simeon Bird}
\affiliation{Department of Physics \& Astronomy, University of California, Riverside, 900 University Ave., Riverside, CA 92521, USA}

\author{Yanhui Yang}
\affiliation{Department of Physics \& Astronomy, University of California, Riverside, 900 University Ave., Riverside, CA 92521, USA}

\author{Rupert Croft}
\affiliation{McWilliams Center for Cosmology, Department of Physics, Carnegie Mellon University, Pittsburgh, PA 15213, USA}

\correspondingauthor{Nianyi Chen}
\email{nianyi.chen7@gmail.com}

\begin{abstract}
Recent pulsar timing array (PTA) observations have detected nanohertz gravitational waves, likely originating from massive black hole binaries (MBHBs). 
The detected amplitude is unexpectedly higher than inferred from the electromagnetic measurements. 
We present new gravitational wave background (GWB) results from the ASTRID simulation. Its large volume and on-the-fly dynamical friction for MBHs provide new insights into the MBHB population, offering a more accurate assessment of its contribution to the observed GWB.
ASTRID predicts a GWB from MBHBs of $h_c=2.8\times10^{-15}$, or $\sim45\%$ of the observed amplitude at $\sim 4\,{\rm nHz}$ and $h_c=2.5\times10^{-16}$ ($5\%$) with $h_c\propto f^{-1.6}$ at $\sim 30\,{\rm nHz}$. These predictions remain below current PTA constraints but align with previous empirical models based on the observed MBH mass functions.
By comparison, TNG300 with post-processed MBH dynamics yields a range between $70-90\%$ ($20\% - 30\%$) of the observed levels at low (high) frequencies. 
At low frequencies, ASTRID predicts that the bulk of the GWB originates from MBHB with masses $M_{\rm tot}=1-3\times 10^9\,M_\odot$ peaking at $z\approx 0.3$, consistent with TNG300. 
Notably, both simulations predict significant GWB contribution from minor mergers ($q<0.2$) by up to $\sim 40\%$.
By tracing the full merger trees of local MBHs in ASTRID, we show that they generate GWs at $\sim 10\%-80\%$ of the maximum signal assuming no accretion and recent equal-mass mergers. 
Finally, we demonstrate the importance of on-the-fly MBH dynamics, the lack of which leads to $3- 5$ times excessive mass growth by merger, and a similar boost to the GWB prediction.


\end{abstract}


\section{Introduction}
\label{section:introduction}


Pulsar timing arrays (PTAs) recently detected the stochastic gravitational wave background (GWB), opening up a new window on Massive Black Holes (MBHs) via their gravitational wave emission. In particular, the nanohertz frequency band contains a stochastic gravitational wave background derived from the sum of all massive black hole binaries on bound Keplerian orbits. A nanohertz GWB has been detected with statistical significance between 2-4 $\sigma$ \citep[e.g.][]{EPTA2023A&A...678A..50E, PPTA2023, NG15-evidence,NG15-binary, CPTA2023,IPTA2024}.
However, the amplitude of the measured GWB is apparently larger than many theoretical predictions. 
In addition, when modeled as a single power law, the slope of the observed GWB appears to be slightly flatter than the $S \propto f^{-13/3}$ expected in the circular, GW-driven case. 

Theoretical predictions of the GWB typically use semi-analytical models (SAMs) to forward model the MBH binary populations from analytical prescriptions of the galaxy mass function, galaxy-MBH connection, merger rate, and/or MBH merging timescales, sometimes combined with simulated galaxy/halo merger trees \citep[e.g.][]{Wyithe2003, Sesana2008, McWilliams2014, Kulier2015ApJ...799..178K, Bonetti2018, ChenSiyuan2019, Simon2023ApJ...949L..24S}.
Other approaches try to bypass these detailed models and estimate a maximum GWB amplitude assuming that all GWB sources come from parents of the $z=0$ MBHs sitting on the local scaling relation \citep[e.g.][]{Sato-Polito2024}.
Recently, more sophisticated predictions have been derived using black hole merger catalogs directly from hydrodynamical simulations with on-the-fly MBH/AGN evolution \citep[e.g.][]{Kelley2017a, Kelley2017b, Siwek2020MNRAS.498..537S}.
However, these simulations do not track MBH dynamics and instead rely on repositioning MBHs to the host galaxy centers, which truncates the MBH merger timescales.
In this work, we present a substantially substantially improved model for the GWB amplitude and its composition using the MBH merger catalog from the cosmological hydrodynamical simulation \texttt{ASTRID}, which is the first large-scale simulation to include a dynamic friction model from dark matter, stars, and gas, following the MBH mergers down to $<{\rm kpc}$ scale over a wide range of mass and redshift. 
We also show as a comparison set results from the \texttt{TNG300} simulation with post-processed dynamical friction modeling and investigate the effect of on-the-fly dynamical friction delays and BH mass functions.

This paper is organized as follows. In Section \ref{sec:sim}, we introduce the two simulations used for this work, with a focus on the modeling of MBHs. In Section \ref{sec:binaries}, we discuss the binary formation timescales and show the properties of MBH mergers and binaries from simulations.
Section \ref{sec:gwb} shows the GWB predictions from the simulations and provides a detailed analysis of the source demographics.
Finally, in Section \ref{sec:mf_merger}, we connect MBH binaries and GWB to the build-up of the local MBH mass function, by investigating the role of binary formation timescales in both the GWB production and the MBH mass growth histories.

\section{Simulations}
\label{sec:sim}
The main results of this letter rely on the \texttt{ASTRID} cosmological hydrodynamic simulation, now run to $z=0$ \citep{Zhou:2025}. \texttt{ASTRID} contains models for galaxy formation including gas cooling, star formation, metal return, BH seeding, merging and accretion as well as supernova and AGN feedback (including both thermal and kinetic modes).
\texttt{ASTRID} is described fully in the introductory papers \citep[e.g.][]{Bird-Astrid, Ni-Astrid, Chen2022, Chen2023, DiMatteo2023, Ni2024}. Here we briefly introduce the basic parameters and BH modeling.
\texttt{ASTRID} runs in a uniquely large $250\, {\rm cMpc}/h$ box with $2\times 5500^3$ particles, thus a 
dark matter resolution element of mass $M_{\rm DM} = 9.94 \times 10^6\,M_\odot$. 
The gravitational softening length is $\epsilon_{\rm g} = 1.5 \, {\rm ckpc}/h$. 

The large volume of \texttt{ASTRID} enables us to investigate rare pairs of very massive BHs with $M_{\rm BH} >10^{8}\,M_\odot$. In \texttt{ASTRID}, BHs are seeded in halos with $M_{\rm halo} > 5 \times 10^9 \hmsun$ and $M_{\rm *} > 2 \times 10^6 \hmsun$, with seed masses stochastically drawn from a power law between $3\times10^{4} \hmsun$ and $3\times10^{5} \hmsun$. 
When the accretion rate is less than $5\%$ of the Eddington ratio and $M_{\rm BH} \gtrsim 10^{8.5} M_\odot$, the BHs enter a kinetic feedback mode, similar to \citet{Weinberger2017}, but with parameters that make kinetic feedback moderately less aggressive \citep{Ni-Camels}. 

Uniquely among large-scale cosmological simulations, \texttt{ASTRID} includes a subgrid dynamical friction model \citep{Tremmel2015,Chen2021} from dark matter, stars, and gas, yielding physically consistent black hole trajectories and velocities.
This model, compared with the repositioning of MBHs used in, e.g.~\texttt{TNG300}, avoids spurious BH mergers and resolves the dynamical friction timescale down to the numerical resolution limit. We will show that including dynamical friction significantly affects the BH binary population and predictions for the GWB.
Two black holes merge when their separation is within two times the gravitational softening length $2\epsilon_g=3\,{\rm ckpc}/h$ and if their kinetic energy is dissipated by dynamical friction and they are gravitationally bound to the local potential.

For a comparison to a simulation using the repositioning model, we employ the \texttt{TNG300-1} (\texttt{TNG300} hereafter) simulation \citep{Pillepich2018, Springel2018, Naiman2018, Nelson2018, Marinacci2018}, with MBH catalog obtained from the data release \citep[][]{Blecha2016MNRAS.456..961B, Kelley2017a}.
\texttt{TNG300} has a box size of $205\,{\rm cMpc}/h$ with $2\times 2500^3$ particles.
In \texttt{TNG300}, BHs have a fixed seed mass $M_{\rm seed}=1.18\times 10^6 M_\odot$ and are seeded in halos with $M_{\rm halo} > 7.38 \times 10^{10} M_\odot$. 
\texttt{ASTRID} and \texttt{TNG300} have similar BH feedback and accretion models. However, in \texttt{TNG300} the BH kinetic feedback mode is enabled when the accretion rate is less than $10\%$ of the Eddington ratio and $M_{\rm BH} \sim 10^{8} M_\odot$, a lower mass threshold than in \texttt{ASTRID}.
In \texttt{TNG300}, BHs are continuously repositioned to the potential minimum of the host halo, and merge with any other BH coming within twice the gravitational softening length.
This can lead to BHs merging a few Gyrs early as well as produce spurious mergers leading to multiple seeding in a single galaxy \citep{Chen2021}.
By construction only central BHs are included in the total MBH mass density in \texttt{TNG300}, and the  possibility of BHs leaving the host galaxy or being ``off-center'' is neglected.

\section{Binary Population from Simulations}
\label{sec:binaries}
\subsection{Simulated Mergers and PTA-Band Binaries}
\label{subsec:catalog}

\begin{figure*}
\centering
\includegraphics[width=0.99\textwidth]{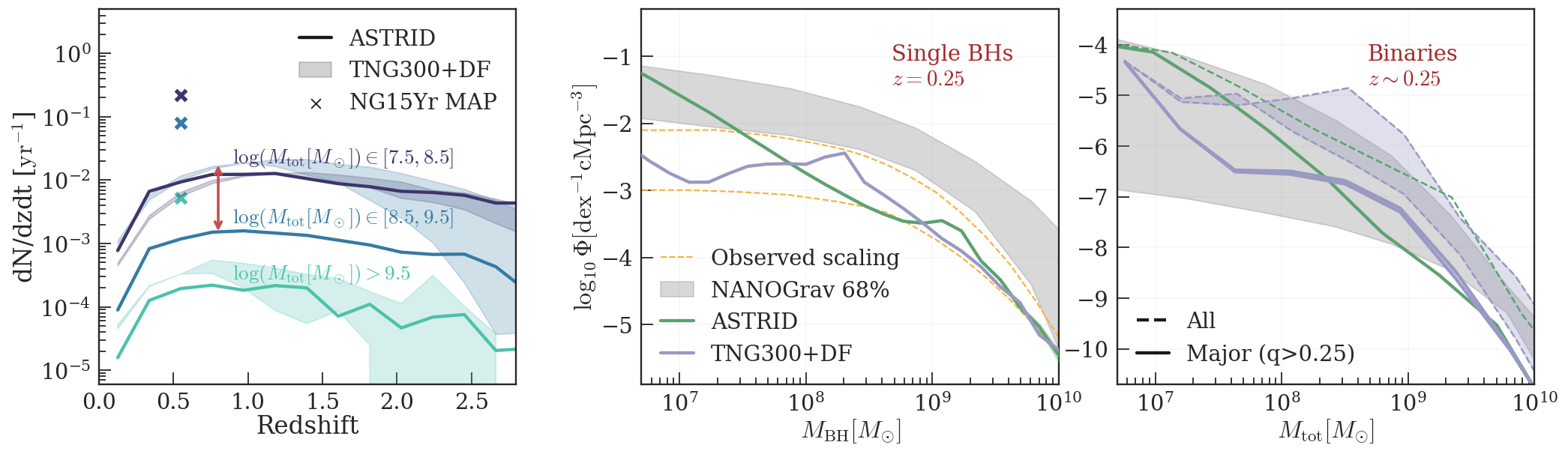}
  \caption{\textit{Left}: The merger rates of major mergers ($q>0.25$) from simulations in mass bins most relevant for PTA-band sources.
  The \textit{solid curves} show the merger rates from \texttt{ASTRID}. The \textit{shaded bands} encloses merger rates from \texttt{TNG300}  with a range of dynamical friction delays. We also include the best-fit binary population from \citetalias{NG15-binary} (\textit{crosses}) at $z=0.5$ for comparison. Simulations show consistent merger rates except for the $10^{8.5}\,M_\odot < M_{\rm ror} < 10^{9.5}\,M_\odot$ mass bin (highlighted by a \textit{red arrow}). \textit{Middle and Right:} Single and binary MBH mass function from \texttt{ASTRID} (\textit{green}) and \texttt{TNG300} (\textit{purple}). We also plot the constraints on single MBH masses from local scaling relations shown in \citet{Sato-Polito2024}(\textit{yellow}) and the constraints from \citetalias{NG15-binary} (\textit{grey shaded}).
  }
  \label{fig:mrate_mtot}
\end{figure*}

While the spatial resolution of large-volume cosmological simulations like \texttt{ASTRID} and \texttt{TNG300} is $\sim{\rm kpc}$, the binary separations corresponding to the PTA frequencies are below the parsec scale.
For \texttt{TNG300}, due to the lack of a dynamical friction phase from the repositioning of BHs, the orbital size at the merger of two BHs is actually larger than the spatial resolution and close to the radius of the host galaxies (i.e. BHs are merged at the first close encounter of their host galaxies).
Hence, we first use analytical estimates of the dynamical friction timescales from \cite{Binney2008} and \cite{Dosopoulou2017} to evolve the \texttt{TNG300} mergers to $\sim{\rm kpc}$ scale pairs (a similar orbital size as \texttt{ASTRID} mergers). 
After that, we apply the same phenomenological binary evolution model as in the NANOGrav 15 Yr MBH binary population analysis \citep[][\citetalias{NG15-binary} hereafter]{NG15-binary} to follow the MBH pairs from kpc separation into the PTA band in both simulations.
We refer readers to Appendix \ref{app:timescale} for a detailed description of the dynamical friction time and binary evolution models.
In particular, we note that the dynamical timescale estimated from the analytical models is consistent with the range of dynamical friction time measured directly from \texttt{ASTRID}.
Here we proceed to showing the simulation mergers and the resulting binary populations after the binary evolution.

The left panel of Figure \ref{fig:mrate_mtot} shows the simulation merger rates of merger mergers with mass ratio $q=M_2/M_1>0.25$ (DF-delayed for \texttt{TNG300}) in mass bins most relevant for the PTA-band GWB.
The merger rates are computed as the number of mergers expected to reach the Earth per year in each redshift interval.
The most noticeable difference between the two simulations comes from major mergers in the mass range $8.5<\log(M_{\rm tot/M_\odot})< 9.5$.
\texttt{TNG300} produces $\sim 8$ times more mergers than \texttt{ASTRID} in this single mass bin.
The number density in all other mass bins differ by at most a factor of $\sim 2$.

The merger rate difference in the $8.5<\log(M_{\rm tot}/M_\odot)< 9.5$ mass bin results from the difference in the single MBH mass functions.
In the middle panel of Figure \ref{fig:mrate_mtot}, we compare the MBH mass functions from two simulations at $z=0.25$, where (as we will show later) the bulk of the GWB is sourced.
We find that the two simulation mass functions are consistent at the most massive end ($M_{\rm BH} > 10^{9.5}\,M_\odot$).
\texttt{ASTRID} has a small bump near $10^9\,M_\odot$, but the biggest difference arises from $M_{\rm BH}<10^9\,M_\odot$.
\texttt{TNG300} BHs grows more efficiently into the $10^8-10^9\,M_\odot$ mass range, leading to 10 times more BHs between $10^8-10^9\,M_\odot$ and a deficit between $10^7-10^8\,M_\odot$ compared to \texttt{ASTRID}.
For reference, we plot $68\%$ confidence intervals of the mass functions inferred from the NANOGrav 15 Yr data \citepalias{NG15-binary}, and the mass function derived from local scaling relations of $M_{\rm BH}$- host galaxy properties derived from electromagnetic (EM) observations \citep[e.g.][]{Sato-Polito2024}.
Both simulations, calibrated to the local MBH and galaxy properties from EM observations, agree with the mass function range covered by local scaling relations.
We note the inconsistency between the single MBH mass function inferred from local scaling relations (and the simulated mass functions) and that inferred from the GWB measurement.
As discussed in previous works such as \cite{Sato-Polito2024}, there is a $2-4.5\sigma$ tension between the maximum GWB based on EM observations and the measured GWB from PTAs, due to the insufficient MBH number density on the most massive end.
Both simulations corroborate this inconsistency and highlight the need for extraordinarily large number of MBHs in the NANOGrav best fit.

The right panel of Figure \ref{fig:mrate_mtot} shows the MBH binary mass function at the same redshift.
After evolving the simulation mergers to the separation (frequency) within the PTA band, we count binaries emitting at $2\,{\rm nHz} < f_{\rm gw, obs} < 40\,{\rm nHz}$ (the frequency range covered by the first 20 frequency bins of the NANOGrav 15-year data) around $z= 0.25$.
The relative abundance of the $q>0.25$ binaries directly reflects the shape of the single MBH mass function (but shifted to the right by $\sim 0.5$ dex because we are plotting the total mass of the binary).
This is because the abundance of major mergers is not sensitive to the dynamical friction models we choose, as all models give relatively short delays in this mass range when $q>0.25$.
The abundance of minor mergers are more sensitive to the estimation of the dynamical friction timescale, and the predictions from \texttt{TNG300} covers a wide range.
The similar shapes of the single and binary mass functions indicate that we can constrain the single MBH population once we have a robust understanding of the binary population, and vice versa.


\section{Stochastic Gravitational Wave Background}
\label{sec:gwb}

\subsection{GWB from \texttt{ASTRID} and \texttt{TNG300}}
\label{subsec:gwb_sim}
With a simulated binary population in the PTA band, we estimate the total characteristic strain expected for each observed frequency bin from the angle- and polarization-averaged strain amplitude of a single binary on circular orbits \citep[e.g.][]{Finn2000}:
\begin{equation}
h_{\mathrm{s}}^2(f_r)=\frac{32}{5 c^8} \frac{(G \mathcal{M})^{10 / 3}}{d_c(z)^2}\left(\pi f_r\right)^{4 / 3}, 
\end{equation}
where $d_c(z)$ is the comoving distance to redshift $z$, and $\mathcal{M}$ is the chirp mass of the binary.
We generate discrete realizations of the binaries in a light-cone from simulations using a weighted sampling method in \texttt{Holodeck} following \cite{Kelley2017b}. 
The signal in each discrete frequency bin at $f$ with a width $\Delta f$ for the $i$th realization is given by
\begin{equation}
\label{eq:discrete}
\left.
    h_{\mathrm{c, i}}^2(f, \Delta f) =  \frac{f}{\Delta f}\sum_j \Lambda_{\rm ij} \, h_{\rm s, j}^2 (f_r) \right|_{f_r= f(1+z)}.
\end{equation}
Here we sum over each simulated binary $j$, and $\Lambda_{ij}$ is a Poisson-distributed weight accounting for the comoving light-cone volume and the binary orbital evolution time.
We describe the detailed formalism and how we compute the weights $\Lambda_{ij}$ in Appendix \ref{app:gwb_calc}.

In Figure \ref{fig:gwb}, we show the characteristic strain prediction from \texttt{ASTRID} and \texttt{TNG300} at the first 15 frequency bins covered by the NANOGrav 15-year measurement \citepalias{NG15-evidence}.
For each simulation, we draw 100 realizations of the binary population to estimate the range of possible strain amplitudes following Equation \ref{eq:discrete}.
We plot the median strain of the 100 realizations, as well as the 68\% interval of the distributions around the median.
For comparison, we show the NANOGrav 15Yr constraints on the GWB in the first 10 frequency bins, and the constraints from the International PTA Collaboration by joining the data from PPTA, NANOGrav, EPTA and InPTA \citep{IPTA2024}.

Both simulations fall short on the GWB from MBH binaries in most frequency bins compared to current observation constraints, especially at high frequencies.
We plot the ratio between the simulation background and the IPTA constraint as a reference in the bottom panel of Figure \ref{fig:gwb}.
The binary population in \texttt{ASTRID} produces $\sim 45\%$ of the measured background in the lowest frequency bin, and this fraction gradually drops to $\sim 5\%$ in the 15th bin ($f\sim 1\,{\rm yr}^{-1}$) with a power-law slope of $h_c \propto f^{-1.6}$ at the high-frequency end.
\texttt{TNG300} produces higher GWB than \texttt{ASTRID} by a factor of $\sim 2-3$, but still lower than the current GWB measurements in most frequency bins.
The drop from a $h_c \propto f^{-2/3}$ scaling is slower in \texttt{TNG300}, with a $h_c \propto f^{-1}$ scaling at $1\,{\rm yr}^{-1}$.
The \texttt{TNG300} band shows uncertainties due to the dynamical friction modeling, with a factor of $\sim 2$ difference between the long-delay and the short-delay models.

Our predictions assume a relatively short binary formation and hardening timescale of $\tau_f=500\,{\rm Myrs}$, which almost maximizes the resulting GWB prediction while retaining a reasonable binary hardening rate.
The absolute maximum of the GWB from the simulations is obtained by assuming (inconsistently) that the MBH pairs immediately enter the PTA band and that the gravitational wave emission is the only energy-loss mechanism driving binary orbital decays.
We show the maximum GWB under this assumption in the bottom panel of Figure \ref{fig:gwb} (ASTRID GW-Only).
The resulting strain amplitude is $\lesssim 10\%$ higher than that of the fiducial hardening model in the two lowest frequency bins and almost identical at higher frequencies, still far short of the NANOGrav observations.
Finally, a nonzero eccentricity can change the shape of the background by shifting the energy emission to higher frequencies and changing the orbital decay rate, but the resulting difference is minor for moderate eccentricity assumptions ($<5\%$ for $e=0.5$, and would only produce lower GWB for $e\gtrsim 0.9$.

\begin{figure}
\centering
\includegraphics[width=0.49\textwidth]{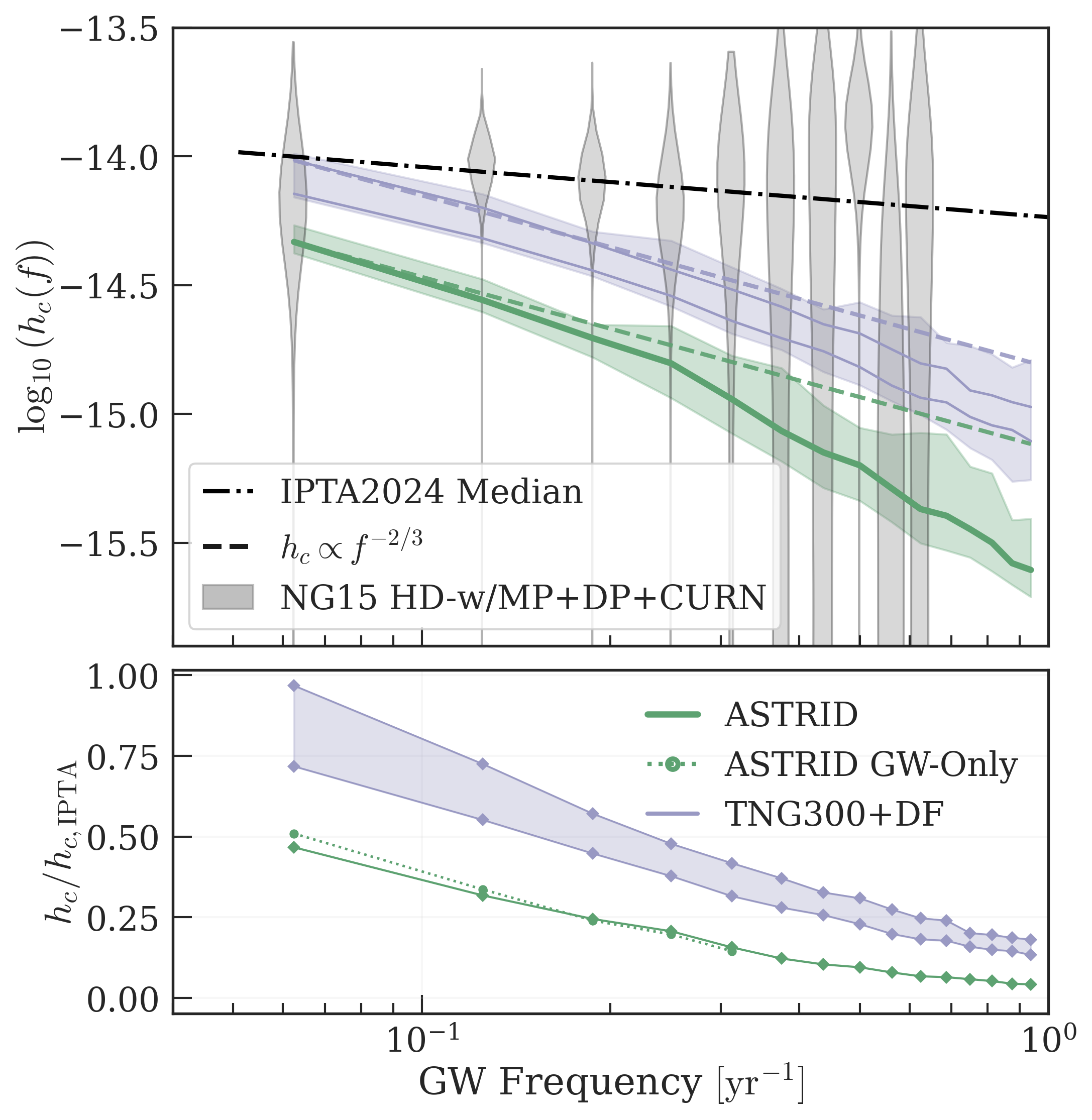}
  \caption{\textit{Top}: The stochastic GWBs generated from the MBH merger population in cosmological simulations \texttt{ASTRID} (\textit{green}) and \texttt{TNG300} (\textit{purple}). 
  We draw 100 realizations of the binary population to estimate the GWB for each simulation.
  We show the median value among all realizations (\textit{solid}) and the $68\%$ intervals (\textit{shaded}).
  We plot the power-law fit to the joint PTA analysis (\textit{black dashed}), and the NANOGrav 15Yr GWB (\textit{grey violins}) for comparison. \textit{Bottom:} The ratio between simulation predicted GWB and the IPTA joint constraints.
  }
  \label{fig:gwb}
\end{figure}

\begin{figure}
\centering
\includegraphics[width=0.48\textwidth]{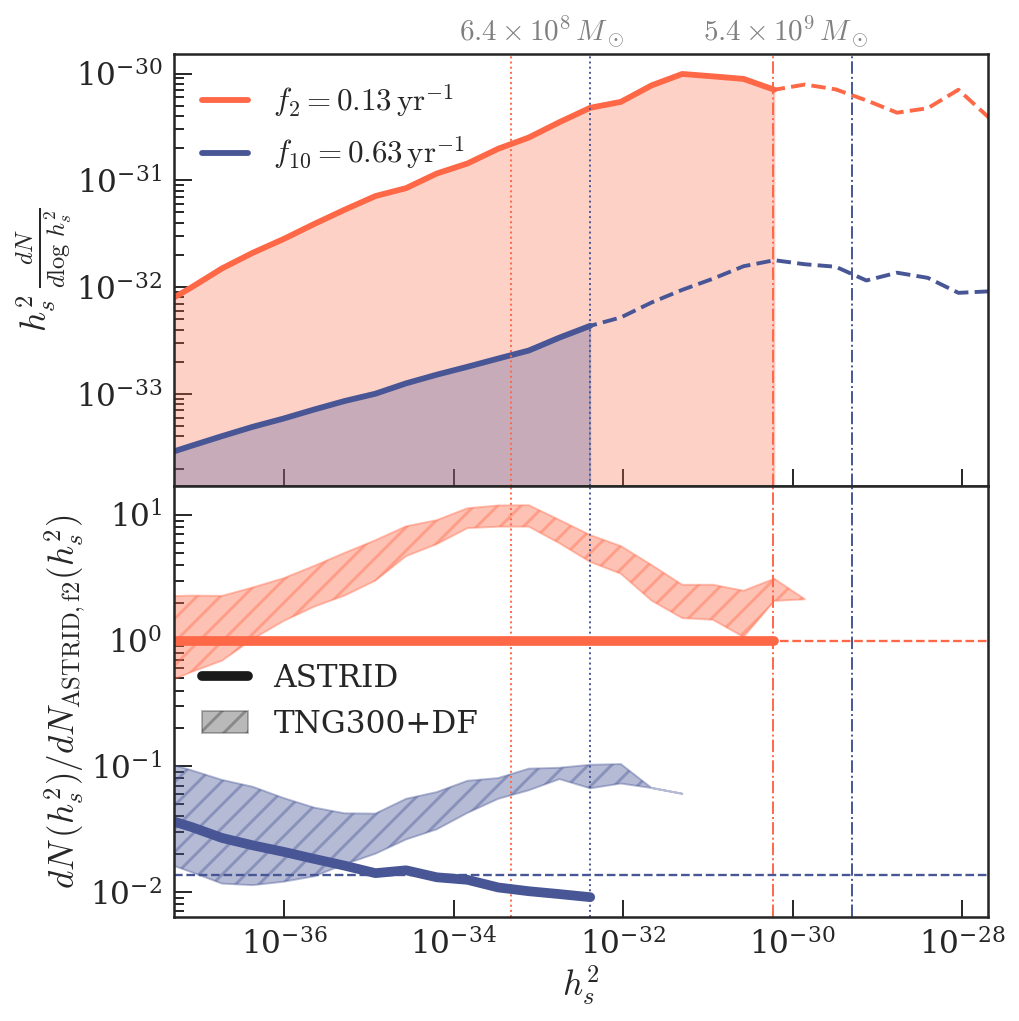}
  \caption{ GW strain distribution of the GWB sources from \texttt{ASTRID} and \texttt{TNG300} simulations. 
  \textit{Top:} The $h_s^2$ distribution weighted by the amplitude in the second ($f_2$, red) and tenth ($f_{10}$, blue) frequency bin of the NG15 data. The solid shaded regions shows the actual strains contributing to the GWB. The dashed curves shows potential foreground sources where the total expected occurrence is 1. For reference, we show in vertical lines the strain corresponding to a $M_{\rm tot}=6.4\times 10^8\,M_\odot$ and a $M_{\rm tot}=5.4\times 10^9\,M_\odot$ equal mass binary at $z=0.3$. \textit{Bottom:} The ratio between the \texttt{ASTRID} high-frequency (\textit{solid blue}), \texttt{TNG300} low (\textit{red shaded}) and high (\textit{blue shaded}) strain distributions to the \texttt{ASTRID} low-frequency distribution.
  }
  \label{fig:hs_distribition}
\end{figure}

The drop in GWB amplitude at high frequencies compared to a power-law scaling is due to the rareness of sources with high strain \citep{Sesana2008}.
For a given MBH binary population, there is a cut-off mass \citep[or strain, as formalized in e.g.][]{Sato-Polito2024b}, above which the total probability of the occurrence of the PTA band is less than one.
Thus we expect only one source above the cutoff strain in any one realization, so that these populations do not contribute to the background.

We illustrate this effect in the top panel of Figure \ref{fig:hs_distribition}, which shows the contribution to the total strain per logarithmic $h_s^2$ bin at two frequencies.
The solid shaded regions are sources actually contributing to the GWB, and the dashed regions are the ``fractional sources'', which exist as individual binaries in the simulations, but whose very efficient orbital evolution results in sampling weights $\Lambda_{ij}\sim 0$ in a light-cone realization.

In \texttt{ASTRID}, the peak of the strain contribution comes from $h_s^2\sim 10^{-31}-10^{-30}$, equivalent to the strain of a $M_{\rm tot}\sim 2\times 10^9\,M_\odot$ equal mass binary at the peak redshift $z\sim 0.3$ (or a higher total mass with lower q).
At both frequencies the cutoff of integer sources takes place at or below the peak of $h_s^2$ contributions.
This effect is more prominent for high frequencies, as loud sources spend less time across this frequency bin.
At $20\,{\rm nHz}$ binaries louder than a $M_{\rm tot}\sim 6\times 10^8\,M_\odot$ major merger do not contribute to the GWB.
The shaded red region under the blue dashed curve is the amount of deviation from a $h_c \propto f^{-2/3}$ scaling in the GWB.

The bottom panel of Figure \ref{fig:hs_distribition} shows the relative $h_s^2$ distribution between the two simulations.
The relative shape of the $h_s^2$ distribution reflects the shapes of the MBH mass function.
Because the $M_{\rm tot}\sim 10^{8.5}\,M_\odot$ ``bump'' in the \texttt{TNG300} mass function coincides with the cutoff strain at the high frequency bin, the \texttt{TNG300} high-frequency background contains louder sources (up to $h_s^2 \sim 10^{-31.5}$) compared to \texttt{ASTRID}, resulting in a shallower slope.
Our results demonstrate that the shape of the GWB can potentially  provide constraints on the MBH mass function beyond current EM observations.

\subsection{GWB Source Demographics}
\label{subsec:gwb_distribution}

\begin{figure*}
  \centering
\includegraphics[width=0.47\textwidth]{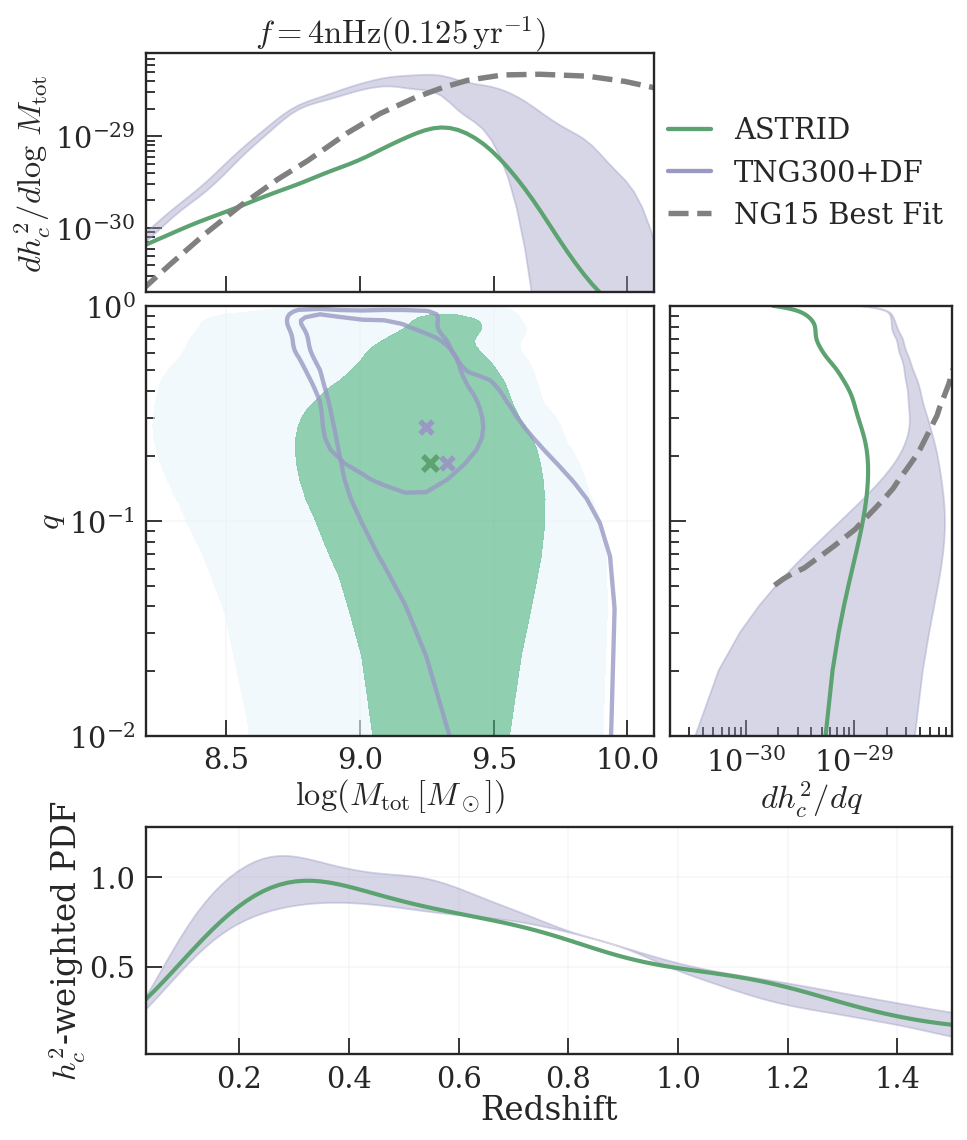}
\includegraphics[width=0.47\textwidth]{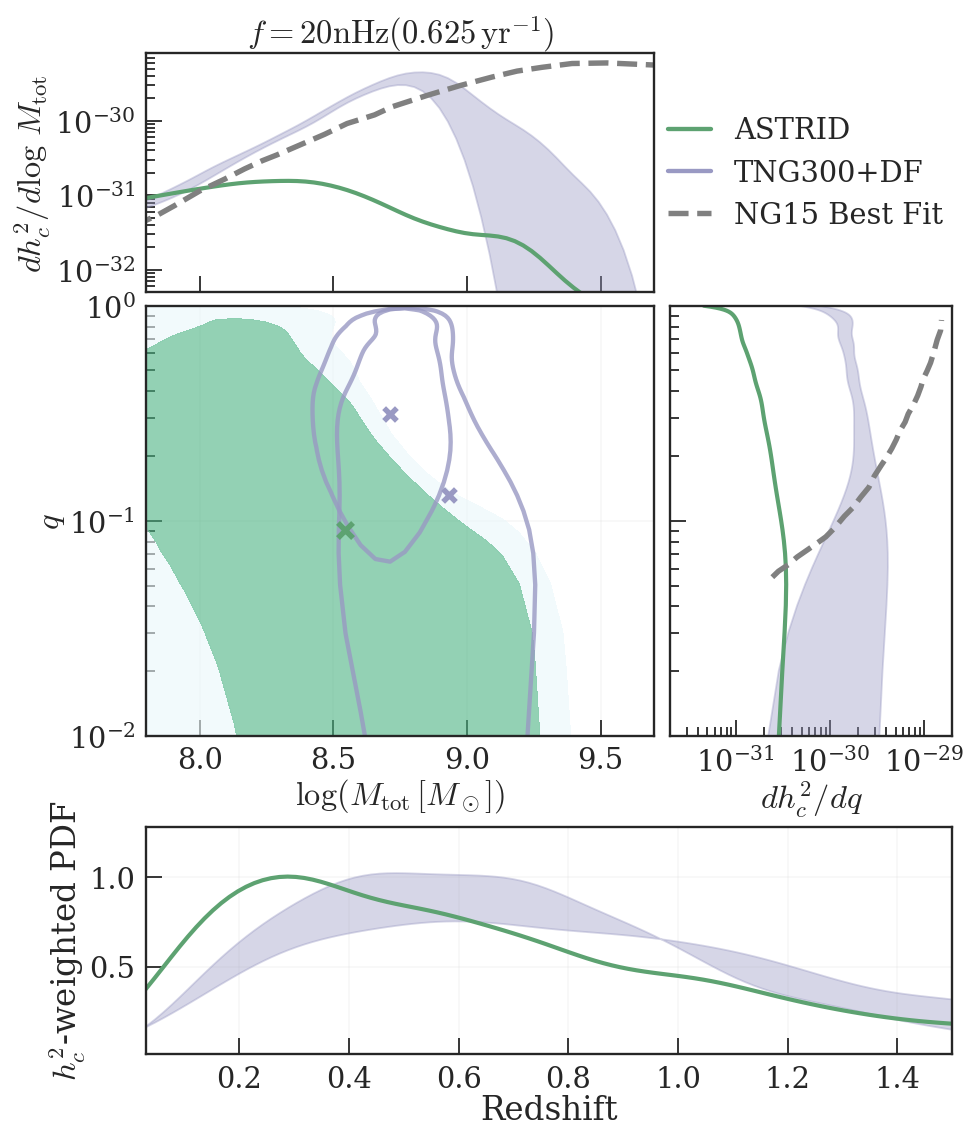}
    \caption{Distribution of total mass, mass ratio and redshift of MBH binaries weighted by their contribution to the GW background ($h_s^2$) at $f=0.125\,{\rm yr}^{-1}$ and $f=0.625\,{\rm yr}^{-1}$. We compare the populations from \texttt{ASTRID} (\textit{green}) and \texttt{TNG300} (\textit{purple}) with the best-fit semi-analytical model from \citetalias{NG15-binary} (\textit{grey dashed}). For the 2D distribution, we plot the peak of each distribution in crosses, and the $68\%$ contribution contour around the peak. For \texttt{ASTRID}, we also plot the $95\%$ contour in light green.}
    \label{fig:mq_distribution}
  \end{figure*}

Now we investigate which binaries contribute to the GWB in each model by analyzing the source distributions, and show where the differences in the GWB amplitude arise.

The top and middle rows of Figure \ref{fig:mq_distribution} shows the source contributions to the GWB at two frequencies, binned by total masses and mass ratios.
The signal produced by any mass/mass ratio range is the area under the 1D curves covered by that range.
The lower left panels in each subplot display the distribution in the 2D plane of $M_{\rm tot}$ and $q$. 
For the 1D distributions we also show the best-fit binary population to the NANOGrav 15Yr data \citepalias{NG15-binary} for comparison.

In the low-frequency bin, the simulations show general agreement on the peak of the source distribution around $M_{\rm tot}=10^9 \sim 3\times 10^9\,M_\odot$ and $q=0.2\sim 0.8$,
and the contribution of the most massive binaries ($M_{\rm tot}=4\times 10^9 \sim 10^{10}\,M_\odot$).
The disagreement arises from the $8.5 <\log(M_{\rm tot}/M_\odot)<9.5$ range, where \texttt{TNG300} binaries produce 10 times more signal.
This is a consequence of the different BH and binary mass functions.
Both simulations predict a relatively flat distribution in $q$, with minor mergers also contributing significantly (up to $\sim 40\%$) to the background.
For the \texttt{TNG300} predictions, the exact level of contribution from minor mergers depends on the assumption of the dynamical friction time, and \texttt{ASTRID} agrees with the short-delay model where minor mergers contribute $\sim 40\%$.
Finally, we note the tension between the simulation predictions and the best-fit model in \citetalias{NG15-binary}.
The \citetalias{NG15-binary} best-fit model favors a significant contribution from very massive BHs ($\log(M_{\rm BH}/M_\odot) > 9.5$), whereas neither simulation produces as many massive binaries in this mass range when fitted with the electromagnetic observations.
The \citetalias{NG15-binary} also inferred a background completely dominated by major mergers with $q\sim 1$, as a result of the particular ansatz of the galaxy pairing chosen in the forward modeling.
Including more contributions from low-q mergers may alleviate the need for a large contribution of very massive BHs to the observed background.

In the high-frequency bin (the right panel of Figure \ref{fig:mq_distribution}), there are more differences between the simulations.
While the GWB in \texttt{ASTRID} consists of sources from a wide range of masses $10^{7.5}-10^{9.3}\,M_\odot$, the \texttt{TNG300} background is dominated by a narrow mass range ($10^{8.5}-10^{9}\,M_\odot$).
The main reason is again the different shape of the BH mass functions.
The dominant sources are shifted to lower-mass binaries compared with the $f=0.12\,{\rm yr}^{-1}$ bin fall within the mass range where the two simulations differ the most.
In the bottom panels of Figure \ref{fig:mq_distribution}, we show the redshift distribution of the GWB sources from the simulations.
The distributions align perfectly for the low-frequency sources ($f=0.12\,{\rm yr}^{-1}$), with a peak at $z=0.25-0.3$.
The high-frequency ($f=0.62\,{\rm yr}^{-1}$) sources are shifted to higher redshift ($z=0.5-0.6$) in \texttt{TNG300} due to the early MBH growth by merger (see the next section), but remain around the same redshift for \texttt{ASTRID} binaries.

\section{Making of the Mass Function by Mergers}
\label{sec:mf_merger}
\begin{figure*}
  \centering
  \includegraphics[width=0.99\textwidth]{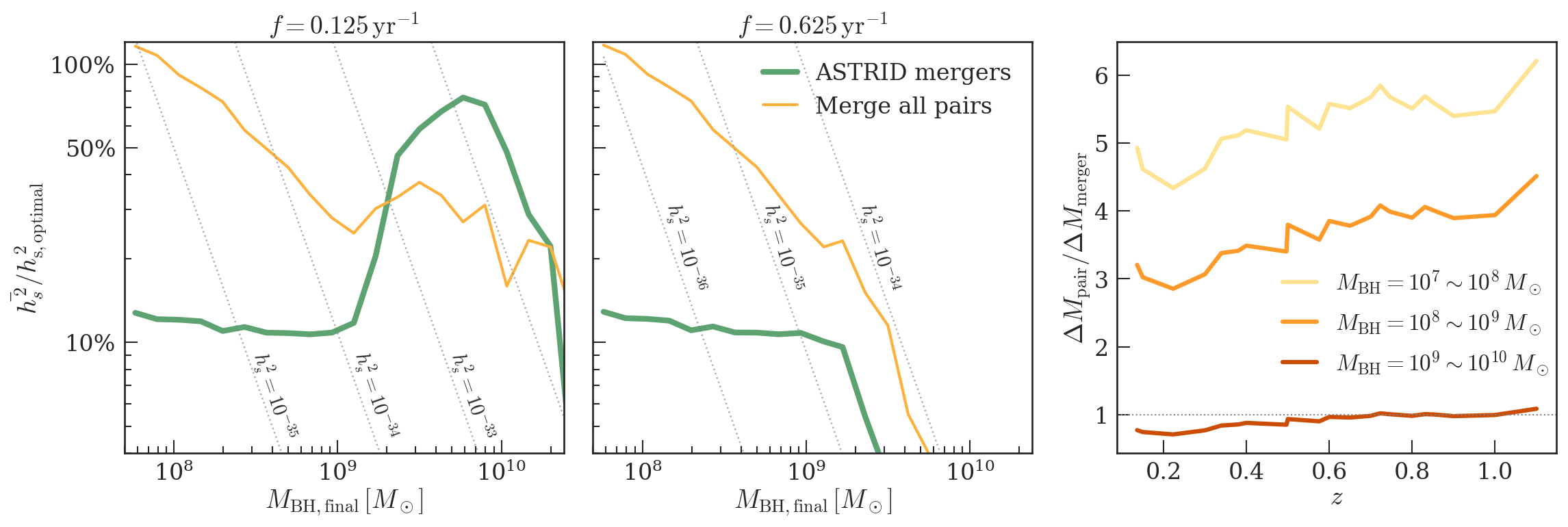}
    \caption{\textit{Left and Middle}: GW production efficiency from past mergers of each local remnant BH as a function of their present-day masses, at $f=0.125\,{\rm yr^{-1}}$ ($4\,{\rm nHz}$, \textit{top}) and at $f=0.625\,{\rm yr^{-1}}$ ($20\,{\rm nHz}$, \textit{bottom}). We characterize this efficiency as a fraction of an optimal strain ($h^2_{\rm s, optimal}$) expected per local BH (see text for details). The prediction from \texttt{ASTRID} merger trees is shown in \textit{green}. We illustrate the effect of merging BHs without a dynamical friction phase using the efficiency computed from fake BH merger trees constructed using MBH pairs separated by $\Delta r=5\,{\rm kpc}$ (\textit{yellow}). \textit{Right}: cumulative mass growth by merger assuming no dynamical friction time compared to the real \texttt{ASTRID} mergers. We show the relative merger mass growth for local remnant BHs in three mass bins. }
    \label{fig:bhmf_strain}
  \end{figure*}

The massive end of the BH mass function at each redshift is closely related to the rate and distribution of mergers prior to that redshift. 
This provides a mapping between the local BH mass function and the expected GWB, with certain assumptions about the merger histories of the local BHs.
In an optimistic scenario of zero accretion and when each local MBH is the descendant of an infinite number of equal-mass mergers, \cite{Sato-Polito2024} compute the maximum GW strain from MBH binaries given a local MBH mass function.

In reality, MBHs have gone through many e-folds of mass growth before entering into the merger-dominated growth phase.
Moreover, efficient binary formation and merger may only take place under certain host galaxy conditions and orbital configurations.
To this end, the full accretion and merger history of MBHs from hydrodynamical simulations such as \texttt{ASTRID} can provide constraints on the mapping between the local BH mass function and the GWB signal.

In the left two panels of Figure \ref{fig:bhmf_strain}, we show the mean $h_s^2$ produced by past mergers of local MBHs in each mass bin.
We trace all mergers experienced by the massive progenitor of each local MBH, and sum up the strain contribution from each MBH according to Equation \ref{eq:discrete}.
We then take the mean of the total $h_s^2$ produced per local MBH in each bin of the local MBH mass function.
To better illustrate the GW efficiency, we plot the expected $h_s^2$ per local BH as a fraction of the optimal strain amplitude, computed assuming no gas accretion and an equal-mass merger for each local BH at $z=0.1$.
The absolute magnitude $h_s^2$ can be read off from the labeled dashed lines.

In the low-frequency bin, BHs heavier than $10^9\,M_\odot$ at $z\sim 0$ are relatively efficient in generating the GWB, making up to $80\%$ of the optimal signal per BH.
This means that most BHs in this bin have gone through a major merger in the recent past, after the phase of significant mass growth by accretion.
However, we see from the bottom panel that this part of the mass function does not contribute to the signal at higher frequencies, because their orbital decay is too rapid at small separations and so the expected occurrence rate is low.
For BHs below $10^9\,M_\odot$, the GW production efficiency drops significantly and stays at a constant value of $\sim 10\%$ the optimal efficiency.
This is because smaller MBHs gain most of their mass through accretion, and sustained a relatively high accretion rate after their last major mergers.
This transition is also related to the BH accretion/feedback mode in \texttt{ASTRID}: kinetic feedback mode starts at above $\sim 10^9\,M_\odot$ in the \texttt{ASTRID} model.
We also find that in \texttt{ASTRID}, the total GW signal produced by the entire merger history is dominated by one event, and this holds for all mass bins.

Finally, we revisit the difference in the MBH mass function between \texttt{ASTRID} and \texttt{TNG300} in the $10^{8}-10^{9}\,M_\odot$ mass bin, since we have shown in previous sections that this discrepancy drives the difference in the amplitudes and slopes of the GWB background predictions.
In particular, we investigate how early mergers of MBH pairs without a dynamical friction phase in \texttt{TNG300} boost mass growth for MBHs in the PTA band.
To do so, we construct fake MBH merger trees built for each local MBH in \texttt{ASTRID}, by assuming that the massive progenitor merged with every MBH that came within $5\,{\rm kpc}$, about twice the gravitational softening in \texttt{TNG300} and the typical separation of MBHs after galaxy mergers.

We first show the effect of early merging on the GW production efficiency in the left two panels of Figure \ref{fig:bhmf_strain}.
We find there is a boost in the GW efficiency by a factor of $5-10$ for BHs with $M_{\rm BH}<10^9\,M_\odot$, and that boost becomes more prominent for smaller BHs.
For the most massive BHs, the total strain per remnant BH decreases with the early mergers.
This is because in \texttt{ASTRID}, these massive BHs had their most recent mergers after they complete the bulk of mass growth by gas accretion and enter into the low-accretion mode. 
Without a dynamical friction delay, they would experience their final major merger earlier, so that they may still grow significantly by gas accretion afterwards.

The right panel of Figure \ref{fig:bhmf_strain} more explicitly shows the effect of merging without a DF phase on the mass growth rate by merger.
We compare the cumulative increase of the MBH masses due to mergers in the fake merger trees constructed from MBH pairs, with the real \texttt{ASTRID} MBH mergers.
The excess (or deficit in the most massive bin) of merger mass gain is shown by the ratio between the fake merger accretion and real merger accretion in each mass bins of the final (real) BH masses.
For MBHs with $M_{\rm BH, final} < 10^9\,M_\odot$, early merger of pairs can lead to a $3-5$ times overestimation of merger mass accretion.
This is likely responsible for the accumulation of MBH in $10^8-10^9\,{M_\odot}$ in the \texttt{TNG300} simulation, and the higher GWB signal.

\section{Conclusion and Discussion}
\label{sec:conclusion}

We have examined the stochastic gravitational-wave background  from massive black hole binaries in the nanohertz band, focusing on the improved predictive power of the \texttt{ASTRID} simulation. \texttt{ASTRID} couples large volume and high resolution with an on-the-fly dynamical friction treatment for MBHs, thereby removing degeneracies inherent to previous simulations that rely on repositioning black holes for mergers and fully post-processed orbital decay. We have shown that by explicitly modeling MBH orbits, \texttt{ASTRID} provides  robust predictions for the GWB from MBH mergers, accounting for 
$ \sim$ 45\% of the current PTA-inferred GWB amplitude at $\sim$ 4nHz, and dropping to 
$\sim 5 \%$ at higher frequencies ($\sim $20nHz) with a slope of $h_c\propto f^{-1.6}$. 
This places the ASTRID-based prediction only a factor of two below the amplitude measured by PTAs, making it directly testable with current and upcoming data releases.

We compared these results against \texttt{TNG300}—a similarly large volume cosmological hydrodynamical simulation that implements MBH mergers with repositioning—to highlight how repositioning systematically increases the MBH mass function in the range relevant for the GWB, leading to a higher GWB amplitude. 
Even when dynamical-friction delays are introduced in post-processing, the initial repositioning framework already imprints more total mass growth by mergers, underscoring the fundamental limitations of that approach. 
In contrast, \texttt{ASTRID} self-consistently tracks MBH orbital evolution from large scales, enabling us to set a reliable upper limit on how much MBH mergers can contribute to the background.
Orbital decay models with short delays and a maximum possible GWB do not significantly differ from our predictions.

At low frequencies ($f=0.12\,{\rm yr}^{-1}$), simulations predict very consistent source distributions, with the bulk of the GWB coming from MBH binaries with binary masses $M_{\rm tot}\approx 2\times 10^9\,M_\odot$, mass ratios $q\approx 0.2$ and at redshift $z\approx 0.3$.
We find that minor mergers can contribute up to $40\%$ of the background.
At higher frequencies, the GWB in \texttt{ASTRID} consists of sources from a wide range of masses ($10^{7.5}-10^{9.3}\,M_\odot$), whereas the \texttt{TNG300} background is dominated by a narrow mass range ($10^{8.5}-10^9\,M_\odot$). 
The binary mass function directly reflects the shape of the BH mass function, the amplitude and slope of the GWB can provide direct constraint on the detailed shape of the BH mass function.

By tracing the full merger trees of local MBHs, we show that they generate GWs at $\sim 10\%-80\%$ of the maximum efficiency computed by assuming no accretion and equal-mass mergers.
Finally, we emphasize the importance of on-the-fly BH dynamics modeling, the lack of which can leads to $\sim 5$ times more mass growth by mergers in the $10^{8}-10^9\,M_\odot$ mass range.

Our findings demonstrate the importance of explicit dynamical friction modeling in achieving GWB predictions from simulations with MBHB. The fact that \texttt{ASTRID}’s predicted amplitude sits below but within a factor of a few of the current PTA measurement underscores how close we are to resolving the MBH binary contribution. 
Continued improvements in PTA sensitivity and further refinements to on-the-fly MBH physics will soon allow us to discriminate among different mass functions and binary evolution models and pin down the detailed role of MBH binaries in shaping the observed low-frequency gravitational-wave sky.
If the current discrepancy persists in the future GWB detections, we may need to reconsider the current MBH-galaxy relations \citep[e.g.][]{Liepold2024ApJ...971L..29L}, or incorporate contribution from new physics models \citep[e.g.][]{Caprini2018, Christensen2019}.

\section*{Acknowledgements}
We would like to thank Gabriela Sato-Polito and Matias Zaldarriaga for constructive discussions.
NC acknowledges support from the Schmidt Futures Fund. 
TDM acknowledges funding from NASA ATP 80NSSC20K0519, NSF PHY-2020295, NASA ATP NNX17AK56G, and NASA ATP 80NSSC18K101.
SB acknowledge funding from NASA ATP 80NSSC22K1897.
\astrid~was run on the Frontera facility at the Texas Advanced Computing Center.

\section*{Data Availability}
The code to reproduce the simulation is available at \url{https://github.com/MP-Gadget/MP-Gadget}, and continues to be developed.
The \astrid snapshots are available at \url{https://astrid-portal.psc.edu/}.
The merger catalog is available upon request.

\bibliography{main}{}

\begin{thebibliography}{}
\expandafter\ifx\csname natexlab\endcsname\relax\def\natexlab#1{#1}\fi
\providecommand{\url}[1]{\href{#1}{#1}}
\providecommand{\dodoi}[1]{doi:~\href{http://doi.org/#1}{\nolinkurl{#1}}}
\providecommand{\doeprint}[1]{\href{http://ascl.net/#1}{\nolinkurl{http://ascl.net/#1}}}
\providecommand{\doarXiv}[1]{\href{https://arxiv.org/abs/#1}{\nolinkurl{https://arxiv.org/abs/#1}}}

\bibitem[{{Agazie} {et~al.}(2023{\natexlab{a}}){Agazie}, {Anumarlapudi}, {Archibald}, {Arzoumanian}, {Baker}, {B{\'e}csy}, {Blecha}, {Brazier}, {Brook}, {Burke-Spolaor}, {Burnette}, {Case}, {Charisi}, {Chatterjee}, {Chatziioannou}, {Cheeseboro}, {Chen}, {Cohen}, {Cordes}, {Cornish}, {Crawford}, {Cromartie}, {Crowter}, {Cutler}, {Decesar}, {Degan}, {Demorest}, {Deng}, {Dolch}, {Drachler}, {Ellis}, {Ferrara}, {Fiore}, {Fonseca}, {Freedman}, {Garver-Daniels}, {Gentile}, {Gersbach}, {Glaser}, {Good}, {G{\"u}ltekin}, {Hazboun}, {Hourihane}, {Islo}, {Jennings}, {Johnson}, {Jones}, {Kaiser}, {Kaplan}, {Kelley}, {Kerr}, {Key}, {Klein}, {Laal}, {Lam}, {Lamb}, {Lazio}, {Lewandowska}, {Littenberg}, {Liu}, {Lommen}, {Lorimer}, {Luo}, {Lynch}, {Ma}, {Madison}, {Mattson}, {McEwen}, {McKee}, {McLaughlin}, {McMann}, {Meyers}, {Meyers}, {Mingarelli}, {Mitridate}, {Natarajan}, {Ng}, {Nice}, {Ocker}, {Olum}, {Pennucci}, {Perera}, {Petrov}, {Pol}, {Radovan}, {Ransom}, {Ray}, {Romano}, {Sardesai}, {Schmiedekamp}, {Schmiedekamp},
  {Schmitz}, {Schult}, {Shapiro-Albert}, {Siemens}, {Simon}, {Siwek}, {Stairs}, {Stinebring}, {Stovall}, {Sun}, {Susobhanan}, {Swiggum}, {Taylor}, {Taylor}, {Turner}, {Unal}, {Vallisneri}, {van Haasteren}, {Vigeland}, {Wahl}, {Wang}, {Witt}, {Young}, \& {Nanograv Collaboration}}]{NG15-evidence}
{Agazie}, G., {Anumarlapudi}, A., {Archibald}, A.~M., {et~al.} 2023{\natexlab{a}}, \apjl, 951, L8, \dodoi{10.3847/2041-8213/acdac6}

\bibitem[{{Agazie} {et~al.}(2023{\natexlab{b}}){Agazie}, {Anumarlapudi}, {Archibald}, {Baker}, {B{\'e}csy}, {Blecha}, {Bonilla}, {Brazier}, {Brook}, {Burke-Spolaor}, {Burnette}, {Case}, {Casey-Clyde}, {Charisi}, {Chatterjee}, {Chatziioannou}, {Cheeseboro}, {Chen}, {Cohen}, {Cordes}, {Cornish}, {Crawford}, {Cromartie}, {Crowter}, {Cutler}, {D'Orazio}, {Decesar}, {Degan}, {Demorest}, {Deng}, {Dolch}, {Drachler}, {Ferrara}, {Fiore}, {Fonseca}, {Freedman}, {Gardiner}, {Garver-Daniels}, {Gentile}, {Gersbach}, {Glaser}, {Good}, {G{\"u}ltekin}, {Hazboun}, {Hourihane}, {Islo}, {Jennings}, {Johnson}, {Jones}, {Kaiser}, {Kaplan}, {Kelley}, {Kerr}, {Key}, {Laal}, {Lam}, {Lamb}, {Lazio}, {Lewandowska}, {Littenberg}, {Liu}, {Luo}, {Lynch}, {Ma}, {Madison}, {McEwen}, {McKee}, {McLaughlin}, {McMann}, {Meyers}, {Meyers}, {Mingarelli}, {Mitridate}, {Natarajan}, {Ng}, {Nice}, {Ocker}, {Olum}, {Pennucci}, {Perera}, {Petrov}, {Pol}, {Radovan}, {Ransom}, {Ray}, {Romano}, {Runnoe}, {Sardesai}, {Schmiedekamp}, {Schmiedekamp},
  {Schmitz}, {Schult}, {Shapiro-Albert}, {Siemens}, {Simon}, {Siwek}, {Stairs}, {Stinebring}, {Stovall}, {Sun}, {Susobhanan}, {Swiggum}, {Taylor}, {Taylor}, {Turner}, {Unal}, {Vallisneri}, {Vigeland}, {Wachter}, {Wahl}, {Wang}, {Witt}, {Wright}, {Young}, \& {Nanograv Collaboration}}]{NG15-binary}
---. 2023{\natexlab{b}}, \apjl, 952, L37, \dodoi{10.3847/2041-8213/ace18b}

\bibitem[{{Agazie} {et~al.}(2024){Agazie}, {Antoniadis}, {Anumarlapudi}, {Archibald}, {Arumugam}, {Arumugam}, {Arzoumanian}, {Askew}, {Babak}, {Bagchi}, {Bailes}, {Bak Nielsen}, {Baker}, {Bassa}, {Bathula}, {B{\'e}csy}, {Berthereau}, {Bhat}, {Blecha}, {Bonetti}, {Bortolas}, {Brazier}, {Brook}, {Burgay}, {Burke-Spolaor}, {Burnette}, {Caballero}, {Cameron}, {Case}, {Chalumeau}, {Champion}, {Chanlaridis}, {Charisi}, {Chatterjee}, {Chatziioannou}, {Cheeseboro}, {Chen}, {Chen}, {Cognard}, {Cohen}, {Coles}, {Cordes}, {Cornish}, {Crawford}, {Cromartie}, {Crowter}, {Cury{\l}o}, {Cutler}, {Dai}, {Dandapat}, {Deb}, {DeCesar}, {DeGan}, {Demorest}, {Deng}, {Desai}, {Desvignes}, {Dey}, {Dhanda-Batra}, {Di Marco}, {Dolch}, {Drachler}, {Dwivedi}, {Ellis}, {Falxa}, {Feng}, {Ferdman}, {Ferrara}, {Fiore}, {Fonseca}, {Franchini}, {Freedman}, {Gair}, {Garver-Daniels}, {Gentile}, {Gersbach}, {Glaser}, {Good}, {Goncharov}, {Gopakumar}, {Graikou}, {Griessmeier}, {Guillemot}, {G{\"u}ltekin}, {Guo}, {Gupta}, {Grunthal}, {Hazboun},
  {Hisano}, {Hobbs}, {Hourihane}, {Hu}, {Iraci}, {Islo}, {Izquierdo-Villalba}, {Jang}, {Jawor}, {Janssen}, {Jennings}, {Jessner}, {Johnson}, {Jones}, {Joshi}, {Kaiser}, {Kaplan}, {Kapur}, {Kareem}, {Karuppusamy}, {Keane}, {Keith}, {Kelley}, {Kerr}, {Key}, {Kharbanda}, {Kikunaga}, {Klein}, {Kolhe}, {Kramer}, {Krishnakumar}, {Kulkarni}, {Laal}, {Lackeos}, {Lam}, {Lamb}, {Larsen}, {Lazio}, {Lee}, {Levin}, {Lewandowska}, {Littenberg}, {Liu}, {Liu}, {Liu}, {Lommen}, {Lorimer}, {Lower}, {Luo}, {Luo}, {Lynch}, {Lyne}, {Ma}, {Maan}, {Madison}, {Main}, {Manchester}, {Mandow}, {Mattson}, {McEwen}, {McKee}, {McLaughlin}, {McMann}, {Meyers}, {Meyers}, {Mickaliger}, {Miles}, {Mingarelli}, {Mitridate}, {Natarajan}, {Nathan}, {Ng}, {Nice}, {Ni{\c{t}}u}, {Nobleson}, {Ocker}, {Olum}, {Os{\l}owski}, {Paladi}, {Parthasarathy}, {Pennucci}, {Perera}, {Perrodin}, {Petiteau}, {Petrov}, {Pol}, {Porayko}, {Possenti}, {Prabu}, {Quelquejay Leclere}, {Radovan}, {Rana}, {Ransom}, {Ray}, {Reardon}, {Rogers}, {Romano}, {Russell},
  {Samajdar}, {Sanidas}, {Sardesai}, {Schmiedekamp}, {Schmiedekamp}, {Schmitz}, {Schult}, {Sesana}, {Shaifullah}, {Shannon}, {Shapiro-Albert}, {Siemens}, {Simon}, {Singha}, {Siwek}, {Speri}, {Spiewak}, {Srivastava}, {Stairs}, {Stappers}, {Stinebring}, {Stovall}, {Sun}, {Surnis}, {Susarla}, {Susobhanan}, {Swiggum}, {Takahashi}, {Tarafdar}, {Taylor}, {Taylor}, {Theureau}, {Thrane}, {Thyagarajan}, {Tiburzi}, {Toomey}, {Turner}, {Unal}, {Vallisneri}, {van der Wateren}, {van Haasteren}, {Vecchio}, {Venkatraman Krishnan}, {Verbiest}, {Vigeland}, {Wahl}, {Wang}, {Wang}, {Witt}, {Wang}, {Wang}, {Wayt}, {Wu}, {Young}, {Zhang}, {Zhang}, {Zhu}, {Zic}, \& {International Pulsar Timing Array Collaboration}}]{IPTA2024}
{Agazie}, G., {Antoniadis}, J., {Anumarlapudi}, A., {et~al.} 2024, \apj, 966, 105, \dodoi{10.3847/1538-4357/ad36be}

\bibitem[{{Begelman} {et~al.}(1980){Begelman}, {Blandford}, \& {Rees}}]{Begelman1980}
{Begelman}, M.~C., {Blandford}, R.~D., \& {Rees}, M.~J. 1980, \nat, 287, 307, \dodoi{10.1038/287307a0}

\bibitem[{{Binney} \& {Tremaine}(2008)}]{Binney2008}
{Binney}, J., \& {Tremaine}, S. 2008, {Galactic Dynamics: Second Edition}

\bibitem[{{Bird} {et~al.}(2022){Bird}, {Ni}, {Di Matteo}, {Croft}, {Feng}, \& {Chen}}]{Bird-Astrid}
{Bird}, S., {Ni}, Y., {Di Matteo}, T., {et~al.} 2022, \mnras, 512, 3703, \dodoi{10.1093/mnras/stac648}

\bibitem[{{Blecha} {et~al.}(2016){Blecha}, {Sijacki}, {Kelley}, {Torrey}, {Vogelsberger}, {Nelson}, {Springel}, {Snyder}, \& {Hernquist}}]{Blecha2016MNRAS.456..961B}
{Blecha}, L., {Sijacki}, D., {Kelley}, L.~Z., {et~al.} 2016, \mnras, 456, 961, \dodoi{10.1093/mnras/stv2646}

\bibitem[{{Bonetti} {et~al.}(2018){Bonetti}, {Sesana}, {Barausse}, \& {Haardt}}]{Bonetti2018}
{Bonetti}, M., {Sesana}, A., {Barausse}, E., \& {Haardt}, F. 2018, \mnras, 477, 2599, \dodoi{10.1093/mnras/sty874}

\bibitem[{{Caprini} \& {Figueroa}(2018)}]{Caprini2018}
{Caprini}, C., \& {Figueroa}, D.~G. 2018, Classical and Quantum Gravity, 35, 163001, \dodoi{10.1088/1361-6382/aac608}

\bibitem[{{Chandrasekhar}(1943)}]{Chandrasekhar1943}
{Chandrasekhar}, S. 1943, \apj, 97, 255, \dodoi{10.1086/144517}

\bibitem[{{Chen} {et~al.}(2024){Chen}, {Mukherjee}, {Matteo}, {Ni}, {Bird}, \& {Croft}}]{Chen2024}
{Chen}, N., {Mukherjee}, D., {Matteo}, T.~D., {et~al.} 2024, The Open Journal of Astrophysics, 7, 28, \dodoi{10.33232/001c.116179}

\bibitem[{{Chen} {et~al.}(2022{\natexlab{a}}){Chen}, {Ni}, {Tremmel}, {Di Matteo}, {Bird}, {DeGraf}, \& {Feng}}]{Chen2021}
{Chen}, N., {Ni}, Y., {Tremmel}, M., {et~al.} 2022{\natexlab{a}}, \mnras, 510, 531, \dodoi{10.1093/mnras/stab3411}

\bibitem[{{Chen} {et~al.}(2022{\natexlab{b}}){Chen}, {Ni}, {Holgado}, {Di Matteo}, {Tremmel}, {DeGraf}, {Bird}, {Croft}, \& {Feng}}]{Chen2022}
{Chen}, N., {Ni}, Y., {Holgado}, A.~M., {et~al.} 2022{\natexlab{b}}, \mnras, 514, 2220, \dodoi{10.1093/mnras/stac1432}

\bibitem[{{Chen} {et~al.}(2023){Chen}, {Di Matteo}, {Ni}, {Tremmel}, {DeGraf}, {Shen}, {Holgado}, {Bird}, {Croft}, \& {Feng}}]{Chen2023}
{Chen}, N., {Di Matteo}, T., {Ni}, Y., {et~al.} 2023, \mnras, 522, 1895, \dodoi{10.1093/mnras/stad834}

\bibitem[{{Chen} {et~al.}(2019){Chen}, {Sesana}, \& {Conselice}}]{ChenSiyuan2019}
{Chen}, S., {Sesana}, A., \& {Conselice}, C.~J. 2019, \mnras, 488, 401, \dodoi{10.1093/mnras/stz1722}

\bibitem[{{Christensen}(2019)}]{Christensen2019}
{Christensen}, N. 2019, Reports on Progress in Physics, 82, 016903, \dodoi{10.1088/1361-6633/aae6b5}

\bibitem[{{Di Matteo} {et~al.}(2023){Di Matteo}, {Ni}, {Chen}, {Croft}, {Bird}, {Pacucci}, {Ricarte}, \& {Tremmel}}]{DiMatteo2023}
{Di Matteo}, T., {Ni}, Y., {Chen}, N., {et~al.} 2023, \mnras, 525, 1479, \dodoi{10.1093/mnras/stad2198}

\bibitem[{{Dosopoulou} \& {Antonini}(2017)}]{Dosopoulou2017}
{Dosopoulou}, F., \& {Antonini}, F. 2017, \apj, 840, 31, \dodoi{10.3847/1538-4357/aa6b58}

\bibitem[{{EPTA Collaboration} {et~al.}(2023){EPTA Collaboration}, {InPTA Collaboration}, {Antoniadis}, {Arumugam}, {Arumugam}, {Babak}, {Bagchi}, {Bak Nielsen}, {Bassa}, {Bathula}, {Berthereau}, {Bonetti}, {Bortolas}, {Brook}, {Burgay}, {Caballero}, {Chalumeau}, {Champion}, {Chanlaridis}, {Chen}, {Cognard}, {Dandapat}, {Deb}, {Desai}, {Desvignes}, {Dhanda-Batra}, {Dwivedi}, {Falxa}, {Ferdman}, {Franchini}, {Gair}, {Goncharov}, {Gopakumar}, {Graikou}, {Grie{\ss}meier}, {Guillemot}, {Guo}, {Gupta}, {Hisano}, {Hu}, {Iraci}, {Izquierdo-Villalba}, {Jang}, {Jawor}, {Janssen}, {Jessner}, {Joshi}, {Kareem}, {Karuppusamy}, {Keane}, {Keith}, {Kharbanda}, {Kikunaga}, {Kolhe}, {Kramer}, {Krishnakumar}, {Lackeos}, {Lee}, {Liu}, {Liu}, {Lyne}, {McKee}, {Maan}, {Main}, {Mickaliger}, {Ni{\c{t}}u}, {Nobleson}, {Paladi}, {Parthasarathy}, {Perera}, {Perrodin}, {Petiteau}, {Porayko}, {Possenti}, {Prabu}, {Quelquejay Leclere}, {Rana}, {Samajdar}, {Sanidas}, {Sesana}, {Shaifullah}, {Singha}, {Speri}, {Spiewak}, {Srivastava},
  {Stappers}, {Surnis}, {Susarla}, {Susobhanan}, {Takahashi}, {Tarafdar}, {Theureau}, {Tiburzi}, {van der Wateren}, {Vecchio}, {Venkatraman Krishnan}, {Verbiest}, {Wang}, {Wang}, \& {Wu}}]{EPTA2023A&A...678A..50E}
{EPTA Collaboration}, {InPTA Collaboration}, {Antoniadis}, J., {et~al.} 2023, \aap, 678, A50, \dodoi{10.1051/0004-6361/202346844}

\bibitem[{{Finn} \& {Thorne}(2000)}]{Finn2000}
{Finn}, L.~S., \& {Thorne}, K.~S. 2000, \prd, 62, 124021, \dodoi{10.1103/PhysRevD.62.124021}

\bibitem[{{Haiman} {et~al.}(2009){Haiman}, {Kocsis}, \& {Menou}}]{Haiman2009}
{Haiman}, Z., {Kocsis}, B., \& {Menou}, K. 2009, \apj, 700, 1952, \dodoi{10.1088/0004-637X/700/2/1952}

\bibitem[{{Katz} {et~al.}(2020){Katz}, {Kelley}, {Dosopoulou}, {Berry}, {Blecha}, \& {Larson}}]{Katz2020}
{Katz}, M.~L., {Kelley}, L.~Z., {Dosopoulou}, F., {et~al.} 2020, \mnras, 491, 2301, \dodoi{10.1093/mnras/stz3102}

\bibitem[{{Kelley} {et~al.}(2017{\natexlab{a}}){Kelley}, {Blecha}, \& {Hernquist}}]{Kelley2017a}
{Kelley}, L.~Z., {Blecha}, L., \& {Hernquist}, L. 2017{\natexlab{a}}, \mnras, 464, 3131, \dodoi{10.1093/mnras/stw2452}

\bibitem[{{Kelley} {et~al.}(2017{\natexlab{b}}){Kelley}, {Blecha}, {Hernquist}, {Sesana}, \& {Taylor}}]{Kelley2017b}
{Kelley}, L.~Z., {Blecha}, L., {Hernquist}, L., {Sesana}, A., \& {Taylor}, S.~R. 2017{\natexlab{b}}, \mnras, 471, 4508, \dodoi{10.1093/mnras/stx1638}

\bibitem[{{Kulier} {et~al.}(2015){Kulier}, {Ostriker}, {Natarajan}, {Lackner}, \& {Cen}}]{Kulier2015ApJ...799..178K}
{Kulier}, A., {Ostriker}, J.~P., {Natarajan}, P., {Lackner}, C.~N., \& {Cen}, R. 2015, \apj, 799, 178, \dodoi{10.1088/0004-637X/799/2/178}

\bibitem[{{Li} {et~al.}(2024){Li}, {Volonteri}, {Dubois}, {Beckmann}, \& {Trebitsch}}]{Li2024}
{Li}, K., {Volonteri}, M., {Dubois}, Y., {Beckmann}, R., \& {Trebitsch}, M. 2024, arXiv e-prints, arXiv:2410.07856, \dodoi{10.48550/arXiv.2410.07856}

\bibitem[{{Liepold} \& {Ma}(2024)}]{Liepold2024ApJ...971L..29L}
{Liepold}, E.~R., \& {Ma}, C.-P. 2024, \apjl, 971, L29, \dodoi{10.3847/2041-8213/ad66b8}

\bibitem[{{Marinacci} {et~al.}(2018){Marinacci}, {Vogelsberger}, {Pakmor}, {Torrey}, {Springel}, {Hernquist}, {Nelson}, {Weinberger}, {Pillepich}, {Naiman}, \& {Genel}}]{Marinacci2018}
{Marinacci}, F., {Vogelsberger}, M., {Pakmor}, R., {et~al.} 2018, \mnras, 480, 5113, \dodoi{10.1093/mnras/sty2206}

\bibitem[{{McWilliams} {et~al.}(2014){McWilliams}, {Ostriker}, \& {Pretorius}}]{McWilliams2014}
{McWilliams}, S.~T., {Ostriker}, J.~P., \& {Pretorius}, F. 2014, \apj, 789, 156, \dodoi{10.1088/0004-637X/789/2/156}

\bibitem[{{Naiman} {et~al.}(2018){Naiman}, {Pillepich}, {Springel}, {Ramirez-Ruiz}, {Torrey}, {Vogelsberger}, {Pakmor}, {Nelson}, {Marinacci}, {Hernquist}, {Weinberger}, \& {Genel}}]{Naiman2018}
{Naiman}, J.~P., {Pillepich}, A., {Springel}, V., {et~al.} 2018, \mnras, 477, 1206, \dodoi{10.1093/mnras/sty618}

\bibitem[{{Nelson} {et~al.}(2018){Nelson}, {Pillepich}, {Springel}, {Weinberger}, {Hernquist}, {Pakmor}, {Genel}, {Torrey}, {Vogelsberger}, {Kauffmann}, {Marinacci}, \& {Naiman}}]{Nelson2018}
{Nelson}, D., {Pillepich}, A., {Springel}, V., {et~al.} 2018, \mnras, 475, 624, \dodoi{10.1093/mnras/stx3040}

\bibitem[{{Ni} {et~al.}(2024){Ni}, {Chen}, {Zhou}, {Park}, {Yang}, {DiMatteo}, {Bird}, \& {Croft}}]{Ni2024}
{Ni}, Y., {Chen}, N., {Zhou}, Y., {et~al.} 2024, arXiv e-prints, arXiv:2409.10666, \dodoi{10.48550/arXiv.2409.10666}

\bibitem[{{Ni} {et~al.}(2022){Ni}, {Di Matteo}, {Bird}, {Croft}, {Feng}, {Chen}, {Tremmel}, {DeGraf}, \& {Li}}]{Ni-Astrid}
{Ni}, Y., {Di Matteo}, T., {Bird}, S., {et~al.} 2022, \mnras, 513, 670, \dodoi{10.1093/mnras/stac351}

\bibitem[{{Ni} {et~al.}(2023){Ni}, {Genel}, {Angl{\'e}s-Alc{\'a}zar}, {Villaescusa-Navarro}, {Jo}, {Bird}, {Di Matteo}, {Croft}, {Chen}, {de Santi}, {Gebhardt}, {Shao}, {Pandey}, {Hernquist}, \& {Dave}}]{Ni-Camels}
{Ni}, Y., {Genel}, S., {Angl{\'e}s-Alc{\'a}zar}, D., {et~al.} 2023, \apj, 959, 136, \dodoi{10.3847/1538-4357/ad022a}

\bibitem[{{Peters}(1964)}]{Peters1964}
{Peters}, P.~C. 1964, Physical Review, 136, 1224, \dodoi{10.1103/PhysRev.136.B1224}

\bibitem[{{Pillepich} {et~al.}(2018){Pillepich}, {Nelson}, {Hernquist}, {Springel}, {Pakmor}, {Torrey}, {Weinberger}, {Genel}, {Naiman}, {Marinacci}, \& {Vogelsberger}}]{Pillepich2018}
{Pillepich}, A., {Nelson}, D., {Hernquist}, L., {et~al.} 2018, \mnras, 475, 648, \dodoi{10.1093/mnras/stx3112}

\bibitem[{{Reardon} {et~al.}(2023){Reardon}, {Zic}, {Shannon}, {Hobbs}, {Bailes}, {Di Marco}, {Kapur}, {Rogers}, {Thrane}, {Askew}, {Bhat}, {Cameron}, {Cury{\l}o}, {Coles}, {Dai}, {Goncharov}, {Kerr}, {Kulkarni}, {Levin}, {Lower}, {Manchester}, {Mandow}, {Miles}, {Nathan}, {Os{\l}owski}, {Russell}, {Spiewak}, {Zhang}, \& {Zhu}}]{PPTA2023}
{Reardon}, D.~J., {Zic}, A., {Shannon}, R.~M., {et~al.} 2023, \apjl, 951, L6, \dodoi{10.3847/2041-8213/acdd02}

\bibitem[{{Sato-Polito} \& {Zaldarriaga}(2024)}]{Sato-Polito2024b}
{Sato-Polito}, G., \& {Zaldarriaga}, M. 2024, arXiv e-prints, arXiv:2406.17010, \dodoi{10.48550/arXiv.2406.17010}

\bibitem[{{Sato-Polito} {et~al.}(2023){Sato-Polito}, {Zaldarriaga}, \& {Quataert}}]{Sato-Polito2024}
{Sato-Polito}, G., {Zaldarriaga}, M., \& {Quataert}, E. 2023, arXiv e-prints, arXiv:2312.06756, \dodoi{10.48550/arXiv.2312.06756}

\bibitem[{{Sesana} {et~al.}(2007){Sesana}, {Haardt}, \& {Madau}}]{Sesana2007b}
{Sesana}, A., {Haardt}, F., \& {Madau}, P. 2007, \apj, 660, 546, \dodoi{10.1086/513016}

\bibitem[{{Sesana} {et~al.}(2008){Sesana}, {Vecchio}, \& {Colacino}}]{Sesana2008}
{Sesana}, A., {Vecchio}, A., \& {Colacino}, C.~N. 2008, \mnras, 390, 192, \dodoi{10.1111/j.1365-2966.2008.13682.x}

\bibitem[{{Simon}(2023)}]{Simon2023ApJ...949L..24S}
{Simon}, J. 2023, \apjl, 949, L24, \dodoi{10.3847/2041-8213/acd18e}

\bibitem[{{Siwek} {et~al.}(2020){Siwek}, {Kelley}, \& {Hernquist}}]{Siwek2020MNRAS.498..537S}
{Siwek}, M.~S., {Kelley}, L.~Z., \& {Hernquist}, L. 2020, \mnras, 498, 537, \dodoi{10.1093/mnras/staa2361}

\bibitem[{{Springel} {et~al.}(2018){Springel}, {Pakmor}, {Pillepich}, {Weinberger}, {Nelson}, {Hernquist}, {Vogelsberger}, {Genel}, {Torrey}, {Marinacci}, \& {Naiman}}]{Springel2018}
{Springel}, V., {Pakmor}, R., {Pillepich}, A., {et~al.} 2018, \mnras, 475, 676, \dodoi{10.1093/mnras/stx3304}

\bibitem[{{Tremmel} {et~al.}(2015){Tremmel}, {Governato}, {Volonteri}, \& {Quinn}}]{Tremmel2015}
{Tremmel}, M., {Governato}, F., {Volonteri}, M., \& {Quinn}, T.~R. 2015, \mnras, 451, 1868, \dodoi{10.1093/mnras/stv1060}

\bibitem[{{Volonteri} {et~al.}(2020){Volonteri}, {Pfister}, {Beckmann}, {Dubois}, {Colpi}, {Conselice}, {Dotti}, {Martin}, {Jackson}, {Kraljic}, {Pichon}, {Trebitsch}, {Yi}, {Devriendt}, \& {Peirani}}]{Volonteri2020}
{Volonteri}, M., {Pfister}, H., {Beckmann}, R.~S., {et~al.} 2020, \mnras, 498, 2219, \dodoi{10.1093/mnras/staa2384}

\bibitem[{{Weinberger} {et~al.}(2017){Weinberger}, {Springel}, {Hernquist}, {Pillepich}, {Marinacci}, {Pakmor}, {Nelson}, {Genel}, {Vogelsberger}, {Naiman}, \& {Torrey}}]{Weinberger2017}
{Weinberger}, R., {Springel}, V., {Hernquist}, L., {et~al.} 2017, \mnras, 465, 3291, \dodoi{10.1093/mnras/stw2944}

\bibitem[{{Wyithe} \& {Loeb}(2003)}]{Wyithe2003}
{Wyithe}, J. S.~B., \& {Loeb}, A. 2003, \apj, 590, 691, \dodoi{10.1086/375187}

\bibitem[{{Xu} {et~al.}(2023){Xu}, {Chen}, {Guo}, {Jiang}, {Wang}, {Xu}, {Xue}, {Nicolas Caballero}, {Yuan}, {Xu}, {Wang}, {Hao}, {Luo}, {Lee}, {Han}, {Jiang}, {Shen}, {Wang}, {Wang}, {Xu}, {Wu}, {Manchester}, {Qian}, {Guan}, {Huang}, {Sun}, \& {Zhu}}]{CPTA2023}
{Xu}, H., {Chen}, S., {Guo}, Y., {et~al.} 2023, Research in Astronomy and Astrophysics, 23, 075024, \dodoi{10.1088/1674-4527/acdfa5}

\bibitem[{Zhou(2025)}]{Zhou:2025}
Zhou, Y. 2025

\end{thebibliography}
\bibliographystyle{aasjournal}

\appendix
\counterwithin{figure}{section}

\section{Binary Evolution Timescales}
\label{app:timescale}

\begin{figure}
  \centering
  \includegraphics[width=0.47\textwidth]{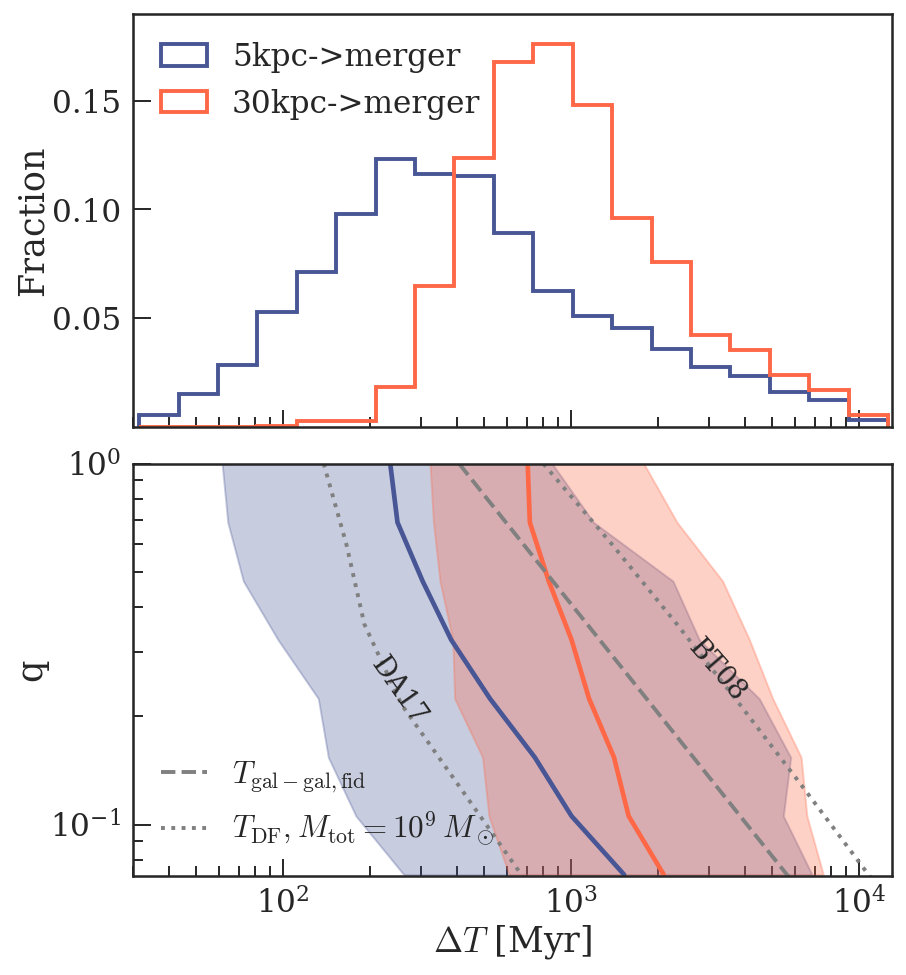}
    \caption{Merging timescale of massive mergers ($M_{\rm tot}>2\times 10^8\,M_\odot$) in \texttt{ASTRID}. \textit{Top:} Distribution of the merging timescale from galaxy pairing ($\Delta r<30{\rm kpc}$) to numerical merger ($\Delta r\sim 1\,{\rm kpc}$ for \texttt{ASTRID}) (red), and from 5kpc to numerical merger (blue). \textit{Bottom:} The MBH merging timescale below $5\,{\rm kpc}$ plotted against the mass ratio $q$. For comparison, we also plotted widely adopted analytical galaxy merging \citep[][]{ChenSiyuan2019} and dynamical friction timescales \citep[][]{Binney2008, Dosopoulou2017}. }
    \label{fig:merging_timesscale}
  \end{figure}

There are several analytical models for estimating the dynamical friction time scales \citep[e.g.][]{Chandrasekhar1943, Dosopoulou2017}, and the results are also sensitive to how one sets the Coulomb logarithm parameter $\Lambda$. 
Meanwhile, these models come with idealized assumptions that may not apply in realistic galaxy mergers \citep[e.g.][]{Chen2024}. 
Therefore, when post-processing \texttt{TNG300} mergers, we aim to encapsulate a range of possible dynamical friction times by taking those that give the shortest and longest timescales while still falling within the range predicted by \texttt{ASTRID}.

In Figure \ref{fig:merging_timesscale}, we show the distribution of the galaxy merger and the dynamical friction timescales for all MBH mergers with $M_{\rm tot} > 5\times 10^8$ and $q > 0.05$ in \texttt{ASTRID}. 
We plot the times between $\Delta r_{\rm pair}=5\,{\rm kpc}$/$30\,{\rm kpc}$ and the \texttt{ASTRID} MBH merger ($\Delta r\sim 1{\rm kpc}$), where $T(\Delta r_{\rm pair})$ is the time that an MBH pair first reaches a separation of $\Delta r_{\rm pair}$.
MBH pairs in this mass range typically spend $1\,{\rm Gyr}$ between $30\,{\rm kpc}$ and $1{\rm kpc}$, and $300-500\,{\rm Myrs}$ between $5\,{\rm kpc}$ and $1{\rm kpc}$.
The timescale has a strong dependency on the mass ratio, as shown in the bottom panel of Figure \ref{fig:merging_timesscale}, and also predicted by analytical dynamical friction models.

In the same plot, we compare the dynamical friction time estimated from \texttt{ASTRID} to a few widely adopted analytical models.
$T_{\rm gal-gal, fid}$ refers to the galaxy merger timescales adopted in semi-analytical models to evolve galaxy pairs at $\sim 5-30\,{\rm kpc}$ into MBH pairs at $a_{\rm init}\sim 1\,{\rm kpc}$ \citep[e.g.][]{ChenSiyuan2019, NG15-binary}. 
More specifically, the line we plot assumes the fiducial parameters in \citetalias{NG15-binary}:
\begin{equation}
T_{\mathrm{gal-gal}}\left(q_{\star}, z^{\prime}\right) =0.5\,{\rm Gyr}\, (1+z)^{-0.5} q_{\star}^{-1},
\end{equation}
where $q_{\star}$ is the galaxy mass ratio.
We note that the full parameterization also allows for a power-law dependency on the primary galaxy mass, but the power-law index is assumed to be $0$ in \citetalias{NG15-binary}.

We also show two estimates of the dynamical friction timescale based on the models from \cite{Binney2008} \citep[applied by e.g.][]{Volonteri2020} and \cite{Dosopoulou2017} \citep[applied by e.g.][]{Katz2020}.
The first model:
\begin{equation}
T_{\mathrm{df, BT08}}=12.5 \frac{1}{\ln\Lambda}
\left(\frac{R_e}{10 \mathrm{kpc}}\right)^2\left(\frac{\sigma}{300 \mathrm{~km} \mathrm{~s}^{-1}}\right)\left(\frac{M_{\mathrm{BH,2}}}{10^8 M_{\odot}}\right)^{-1}  \, \mathrm{Gyr}
\end{equation}
computes the sinking time of a bare MBH sinking in an isotropic potential and provides an upper bound to our estimation. 
$R_e$ is the half-mass radius of the primary galaxy, $\sigma$ is the velocity dispersion of the primary galaxy, $M_{\mathrm{BH,2}}$ is the mass of the secondary MBH, and $\ln\Lambda=1+M_{\mathrm{gal}} / M_{\mathrm{BH}}$. 
We also include a factor of $0.3$ for moderately eccentric orbits.
The second model
\begin{equation}
    T_{\rm df, DA17} = \max \left(T_{\mathrm{df, BT08}}/1000, T_{\star}^{\mathrm{strip}}\right)
\end{equation}
with 
\begin{equation}
T_{\star}^{\mathrm{strip}}=0.3 \frac{1}{\ln \Lambda}\left(\frac{R_e}{10 \mathrm{kpc}}\right)\left(\frac{\sigma}{300 \mathrm{~km} \mathrm{~s}^{-1}}\right)^2\left(\frac{100 \mathrm{~km} \mathrm{~s}^{-1}}{\sigma_s}\right) \mathrm{Gyr}
\end{equation}
assumes the optimistic case where the secondary MBH is embedded in a stellar core $1000$ times its mass and thus sinks $\sim 1000$ times faster. It then takes into account the tidal stripping of the stellar core during infalling with $T_{\star}^{\mathrm{strip}}$. 
Here $\sigma_s$ is the velocity dispersion of the secondary galaxy.

The corresponding timescales for the three models above are shown in the bottom panel of Figure \ref{fig:merging_timesscale}. 
For the two DF models, we show the estimations for an MBH pair with $M_{\rm tot}=10^9\,{M_\odot}$.
The analytical estimations are in broad agreement with the \texttt{ASTRID} prediction (blue and red curves), although there is an order of magnitude difference between the models.
Since realistic galaxy and MBH merger scenarios are often complex and not entirely captured by one of the models (as can be seen also in the wide scatter of the \texttt{ASTRID} prediction), we apply both the DA17 and BT08 models to evolve the \texttt{TNG300} MBH mergers down to $\Delta r\sim 1\,{\rm kpc}$ to bound the \texttt{TNG300} predictions.

Energy loss due to gravitational radiation alone cannot bring most MBH pairs to coalescence within the Hubble time. 
Previous works have shown that multiple channels of orbital delays through interactions with field stars, dark matter, and gas are crucial for the formation of a bound MBH binary and the final coalescence \citep[e.g.][]{Chandrasekhar1943, Begelman1980, Sesana2007b, Haiman2009}.
However, direct modeling of these processes from the environmental information in cosmological simulations is difficult \citep[but see e.g.][for a recent attempt]{Li2024}. Therefore, we use the the phenomenological binary evolution model in the \texttt{holodeck} package to model the environment-driven orbital decay \citep{Kelley2017a, NG15-binary}.
The total orbital decay rate is the superposition of the environment-driven decay and the gravitational-wave driven decay:
\begin{equation}
\label{eq:dadt_tot}
    \left.\frac{d a}{d t}\right|_{\text {tot }} = \left.\frac{d a}{d t}\right|_{\text {gw }} + \left.\frac{d a}{d t}\right|_{\text {phenom }},
\end{equation}
where the gravitational-wave-driven orbital decay rate is given by \citep[][]{Peters1964}:
\begin{equation}
\label{eq:dadt_gw}
\left.\frac{d a}{d t}\right|_{\mathrm{gw}}=-\frac{64 G^3}{5 c^5} \frac{m_1 m_2 M_{\rm tot}}{a^3},
\end{equation}
and the parametrized phenomenological model takes the form
\begin{equation}
\label{eq:dadt_ph}
\left.\frac{d a}{d t}\right|_{\text {phenom }}=H_a \cdot\left(\frac{a}{a_c}\right)^{1-v_{\text {inner }}} \cdot\left(1+\frac{a}{a_c}\right)^{v_{\text {inner }}-v_{\text {outer }}}.
\end{equation}
In this model, $H_a$ is the normalization factor, $\nu_{\rm inner}$ and $\nu_{\rm outer}$ are the two power-law slopes of various binary hardening phases, with the transition at the critical separation $a_c$.
We note that normalization is done by setting the total lifetime of the binary $\tau_f=\int_{a_{\mathrm{init}}}^{a_{\mathrm{isco}}}\left(\frac{d a}{d t}\right)^{-1} d a$, instead of $H_a$, and $a_{\rm init}$ here sets the initial separation from which the phenomenological model starts.

We adopt the same fiducial values for the binary evolution parameters $a_c=100\, {\rm pc}$, $\nu_{\rm outer}=2.5$ as in \citetalias{NG15-binary} for a direct comparison.
For the free parameters, we take the best-fit value of the \textit{Phenom+Astro} analysis in \citetalias{NG15-binary} at $\nu_{\rm inner}=-0.45$ and $\tau_{f}=500\,{\rm Myr}$ (\citetalias{NG15-binary} favors a super efficient binary evolution and we have chosen a more conservative timescale here), and note that this choice pushes the GWB estimate towards the higher end for a given population of MBH pairs.
Finally, $a_{\rm init}$ is set at the MBH separation immediately before merger in \texttt{ASTRID} and thus slightly different between each merger (due to the gravitational-bound check). The values are narrowly distributed around $1\,{\rm kpc}$, which is also the fixed $a_{\rm init}=1\,{\rm kpc}$ assumed in \citetalias{NG15-binary}. This also justifies the choice of the same $\tau_{f}$ for a direct comparison.

\section{Estimating GWB from Simulation Binaries}
\label{app:gwb_calc}
We follow the steps laid out in \cite{Sesana2008} to derive the expression of the total characteristic strain amplitude produced by a population of binaries.
The total observed characteristic strain at frequency $f_{\rm obs}$ produced by the superposition of gravitational radiations from MBH binaries can be computed by integrating over distribution of sources
\begin{equation}
\label{eq:strain_1}
\left.
h_{\mathrm{c}}^2(f)=\int d M_{\rm tot} d q d z \frac{\partial^4 N}{\partial M_{\rm tot} \partial q \partial z \partial \log f_r} h_{\mathrm{s}}^2\left(f_r\right)\right|_{f_r= f(1+z)}.
\end{equation}
Here $dN$ is the number of sources in the differential $M_{\rm tot}$, $q$ and $z$ bins emitting at the source-rest frame log frequency bin [$\log f_r$, $\log f_r + d\log f_r$]. $h_s^2 (f_r)$ is the angle- and polarization-averaged strain amplitude of a single binary on circular orbits, and can be expressed in terms of the source properties as \citep[e.g.][]{Finn2000}:
\begin{equation}
h_{\mathrm{s}}^2(f_r)=\frac{32}{5 c^8} \frac{(G \mathcal{M})^{10 / 3}}{d_c(z)^2}\left(\pi f_r\right)^{4 / 3}, 
\end{equation}
where $d_c(z)$ is the comoving distance to redshift $z$, and $\mathcal{M}$ is the chirp mass of the binary related to the MBH masses ($m_1, m_2$), the total mass $M_{\rm tot}$, and the mass ratio $q$ by
$$
\mathcal{M}=\frac{\left(m_1 m_2\right)^{3 / 5}}{(m_1+m_2)^{1 / 5}}=M_{\rm tot} \frac{q^{3 / 5}}{(1+q)^{6 / 5}}.
$$

We can get a more intuitive expression of Equation \ref{eq:strain_1} in terms of the comoving number density of sources $n=\frac{\partial N}{\partial V_c}$ by noting that
$$
\frac{\partial^4 N}{\partial M_{\rm tot} \partial q \partial z \partial \log f_r}  
= \frac{\partial n}{\partial M_{\rm tot} \partial q \partial z} 
\underbrace{\frac{\partial V_c}{\partial z} \frac{\partial z}{\partial t} }_{V_{\rm fac}}
\underbrace{\frac{\partial t}{\partial \log f_r}}_{t_{\rm fac}}.
$$
Substituting this relation to Equation \ref{eq:strain_1}, we get
\begin{equation}
\label{eq:strain_2}
\left.
 h_{\mathrm{c}}^2(f)
 =\int d M_{\rm tot} d q d z  
 \frac{\partial n}{\partial M_{\rm tot} \partial q \partial z}  
 V_{\rm fac} t_{\rm fac} 
 h_{\mathrm{s}}^2\left(f_r\right) \right|_{f_r= f(1+z)}.
\end{equation}
Here $V_{\rm fac}$ is the volume of the comoving shell of signals reaching the Earth per unit time and can be reduced to a simple expression
$$
V_{\rm fac} = 4 \pi c(1+z) d_c^2.
$$
The time factor $t_{\rm fac}$ represents the rate at which a binary passes through a logarithmic frequency bin.
The longer a source population spends in a frequency bin, the more they would contribute to the observed GWB at that frequency.
We relate it to the rate of binary orbital decay as
$$
t_{\rm fac} = \left(\frac{da}{dt} \right)^{-1}  \frac{da}{df_r} \frac{df_r}{d\log f_r} = -\frac{2}{3} a  \left(\frac{da}{dt} \right)^{-1},
$$
where $a$ is the binary separation, and we have used the $f_{\rm orb}-a$ relation for Keplerian orbits and the relation between GW emission frequency and $f_{r}=2f_{\rm orb}$ for a circular binary in the derivation above.
The total orbital decay rate $\frac{da}{dt}$ is described previously in  Equations \ref{eq:dadt_tot}-\ref{eq:dadt_ph}.

The expression in Equation \ref{eq:strain_2} is convenient because it separates the distribution of the binaries and the observation frequencies.
It can be used to convert any source number density distribution into the number of observed sources at a given frequency, with the weight $V_{\rm fac}\,t_{\rm fac}$.
In the real Universe and in the simulations, the GW sources are discrete, such that the integral of continuous distribution becomes a sum of a discrete terms.
We approximate the source number density $n(z)=\frac{\partial N(z)}{\partial V_c}$ with the number of binaries in the simulation $n(z)=N_{\rm sim}(z)/{V_{\rm sim}}$, and so each simulation binary $j$ has a mean weight of:
\begin{equation}
    \Lambda_j = \frac{V_{\rm fac, j}\, t_{\rm fac, j}}{V_{\rm sim}}
\end{equation}
To create discrete realizations of GWB sources from simulation events, we follow \citet{Kelley2017a} and assume the weight of each simulation binary follows the Poisson distribution $\Lambda_{\rm ij}\sim {\rm Poisson}(\Lambda_j)$, so the estimated stochastic background in realization $i$ is
\begin{equation}
\left.
    h_{\mathrm{c, i}}^2(f) =  \sum_j \Lambda_{\rm ij} \, h_{\rm s, j}^2 (f_r)  \right|_{f_r= f(1+z)}.
\end{equation}
Finally, we also note the discreteness in the observable frequency bins, with a minimum bin size set by the total observation time $\Delta f=1/T_{\rm obs}$. 
A source observed at frequency $f$ stays in the frequency bin $[f, f+\Delta f]$ for $n=f/\Delta f$ cycles. The final signal in each discrete frequency bin for the $i$th realization is given by
\begin{equation}
\label{eq:discrete}
\left.
    h_{\mathrm{c, i}}^2(f, \Delta f) =  \frac{f}{\Delta f}\sum_j \Lambda_{\rm ij} \, h_{\rm s, j}^2 (f_r) \right|_{f_r= f(1+z)}.
\end{equation}
We show results of the GWB and contributing sources estimated through this equation.
\end{document}